\documentclass{llncs}\usepackage[]{graphicx}\usepackage[]{color}
\makeatletter
\def\maxwidth{ %
  \ifdim\Gin@nat@width>\linewidth
    \linewidth
  \else
    \Gin@nat@width
  \fi
}
\makeatother

\definecolor{fgcolor}{rgb}{0.345, 0.345, 0.345}

\usepackage{framed}
\makeatletter
 {\par\unskip\endMakeFramed%
 \at@end@of@kframe}
\makeatother

\definecolor{shadecolor}{rgb}{.97, .97, .97}
\definecolor{messagecolor}{rgb}{0, 0, 0}
\definecolor{warningcolor}{rgb}{1, 0, 1}
\definecolor{errorcolor}{rgb}{1, 0, 0}
\usepackage{alltt} % authordraft
\usepackage{epsfig,endnotes,url}

\usepackage{xspace}

\usepackage[english]{babel}
\usepackage[latin1]{inputenc}
\usepackage{amsmath, amssymb, latexsym} % NOT FOR SPRINGER: amsthm 

\usepackage[ShowComments]{mycomment}

\usepackage{graphicx}
\usepackage[dvipsnames,table]{xcolor}

\usepackage[disable,colorinlistoftodos]{todonotes}   % disable

\newcommand{\pretty}[1]{} %\todo[color=Lavender, inline]{Pretty Printing---#1}}
\newcommand{\fixme}[1]{\todo[color=Red, inline]{\textbf{FIX!!!}---#1}}

\usepackage{multirow}
\usepackage{booktabs}
\usepackage{rotating}

\usepackage{paralist}

\usepackage{balance}

\usepackage{versions}
\excludeversion{DocumentVersionConference}
\includeversion{DocumentVersionTR}
\includeversion{DocumentVersionLNCS}
\excludeversion{RedundantContent}

\usepackage[caption=false]{subfig}

\newtheorem{researchquestion}{RQ}

\newcommand{\vari}[1]{\ensuremath{\mathit{#1}\xspace}}
\newcommand{\const}[1]{\ensuremath{\mathsf{#1}\xspace}}

\newcommand{\CASCAde}{ERC Starting Grant CASCAde (GA n\textsuperscript{o}716980)}
\IfFileExists{upquote.sty}{\usepackage{upquote}}{}
\begin{document}

% DATA IMPORT
\newcommand{\sampleBoxplot}{
\begin{figure}[tb]

\includegraphics[width=\maxwidth]{figure/data_sample_sizes-1} 
\caption{Boxplot of the sample sizes of the SLR sample}
\label{fig:sampleBoxplot}
\end{figure}
}

\newcommand{\sampleTests}{
% latex table generated in R 3.6.2 by xtable 1.8-4 package
% Mon Oct  5 16:30:53 2020
\begin{table}[ht]
\centering
\caption{Sample Refinement on Extracted Effect Sizes} 
\label{tab:sampleTests}
\begingroup\footnotesize
\begin{tabular}{lrr}
  \toprule
\textbf{Phase} & Excluded & Retained \\ 
  \midrule
Total effects extracted & 0 & 650 \\ 
   \midrule
 \quad \textsf{statcheck} automated extraction &  &  252 \\
 \quad Test statistic manual coding &   &  89 \\
 \quad Means \& SD manual coding    &   &  309 \\
 \midrule
 \textit{Refinement in this study}\\
Independent-samples test on dependent sample & 46 & 604 \\ 
  Treated proportion as $t$-distribution & 8 & 596 \\ 
  Reported dependent-samples test w/o correlation & 62 & 534 \\ 
  Reported $\chi^2$ without $\vari{df}$ & 5 & 529 \\ 
  $\chi^2$ with $\vari{df} > 1$ without contingency & 72 & 457 \\ 
  Multi-way $F$-test & 22 & 435 \\ 
  Yielded infinite ES or variance & 3 & 432 \\ 
  Duplicate of other coded test & 1 & 431 \\ 
   \bottomrule
\end{tabular}
\endgroup
\end{table}
}

% KNITR CHILD with R Stats code.

% ----------------------------------------

% ----------------------------------------

\newcommand{\powerSimPlot}{
\begin{figure}[tb]

\includegraphics[width=\maxwidth]{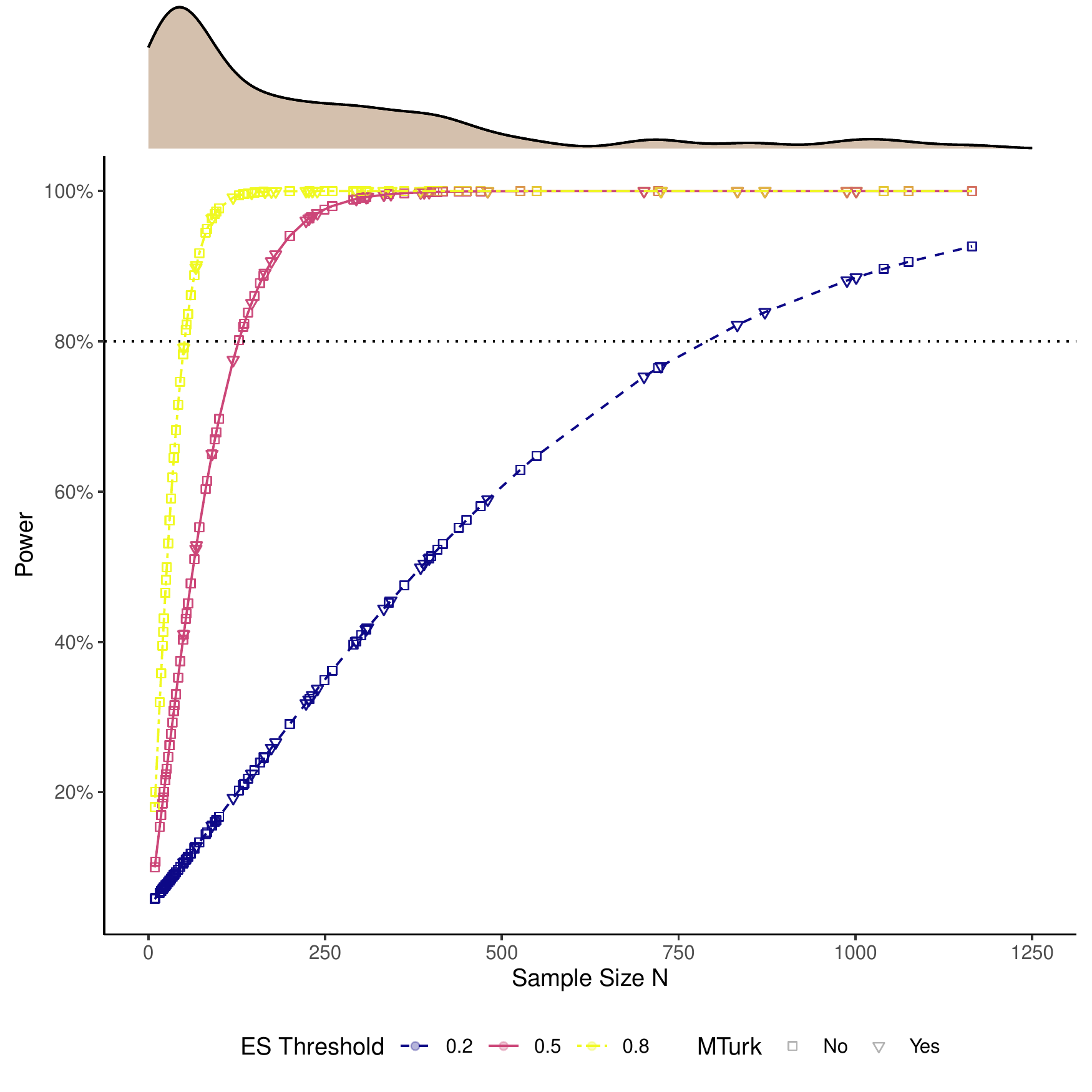} 
\caption{Upper bound of power against SMD effect thresholds and observed study sample sizes $N$ in SLR.}
\label{fig:powerSimPlot}
\end{figure}
}

\newcommand{\powerSimRidgePlot}{
\begin{figure}[tb]

\includegraphics[width=\maxwidth]{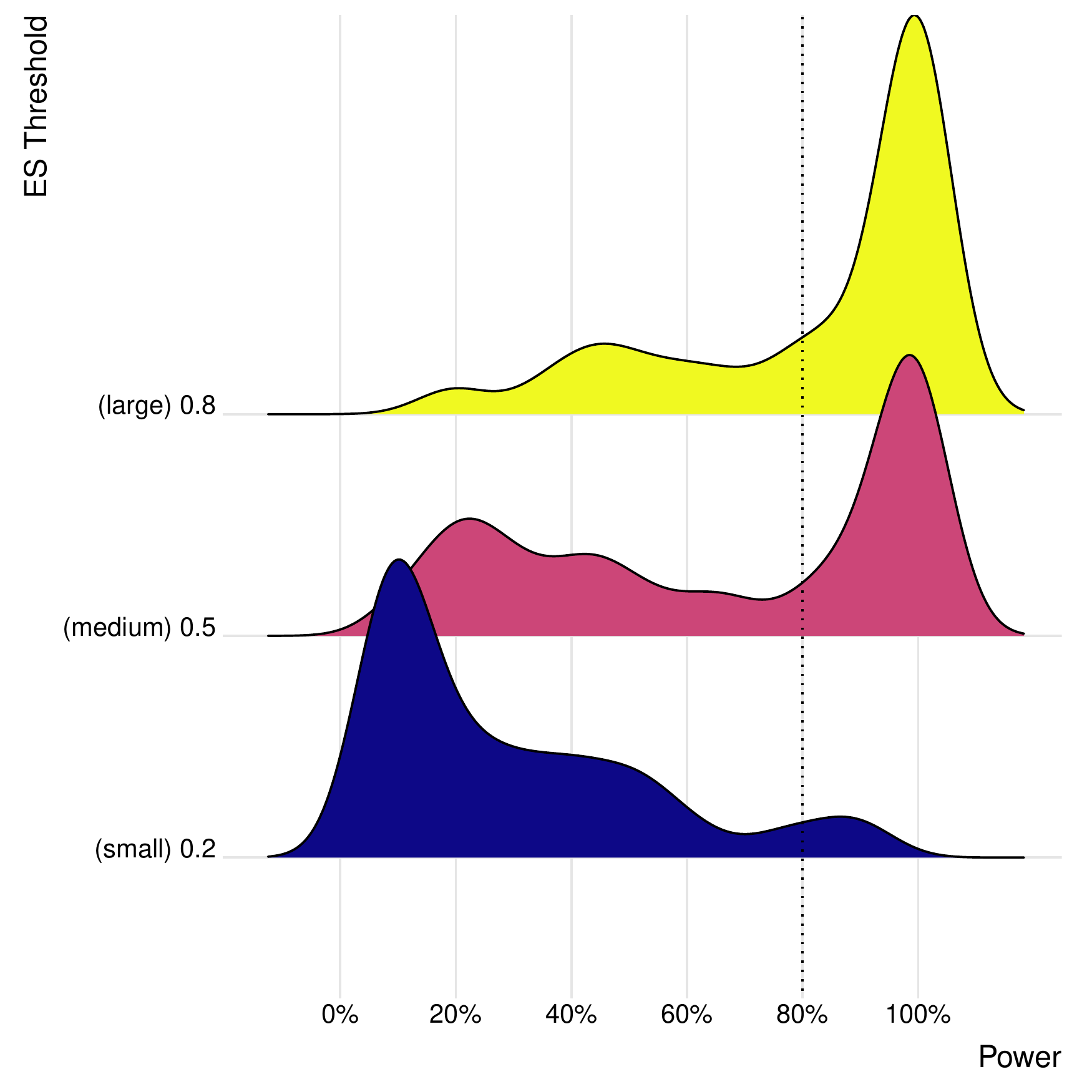} 
\caption{Ridge density plot of upper bound of power against Standardized Mean Difference (SMD) effects and $112$ observed study sample sizes $N$ in SLR.}
\label{fig:powerSimRidgePlot}
\end{figure}
}

\newcommand{\powerSimCombinedUB}{
\begin{figure}[tbp]
\centering\captionsetup{position=bottom}
\begin{minipage}{0.49\textwidth}%
\subfloat[Upper-Bound Power by Sample Size]{
\label{fig:powerSimUB}
\centering
\includegraphics[width=\maxwidth]{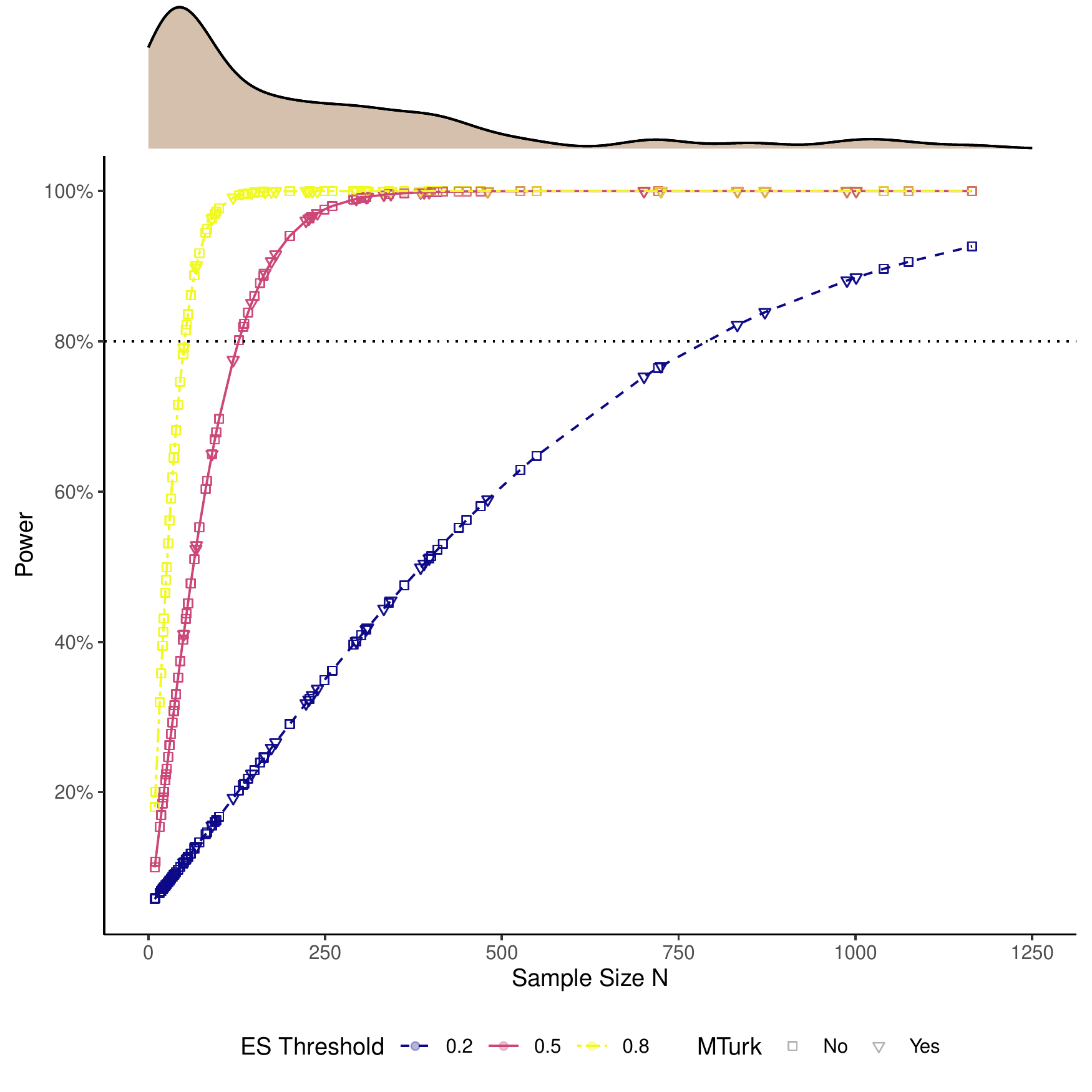} 
}
\end{minipage}~
\begin{minipage}{0.49\textwidth}%
\subfloat[Upper-Bound Power Density]{%
\label{fig:powerSimRidgeUB}
\centering
\includegraphics[width=\maxwidth]{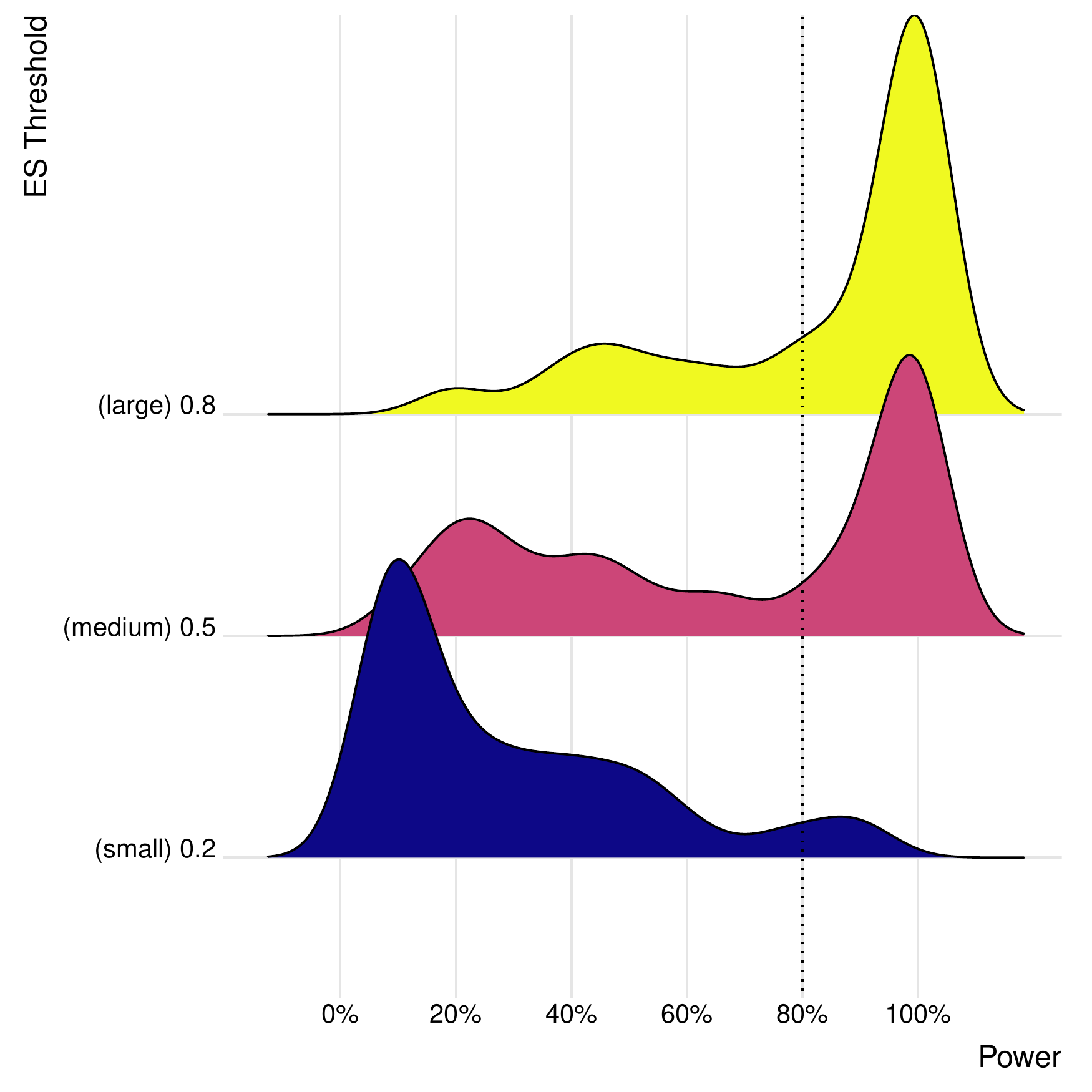} 
}
\end{minipage}
\caption{Upper-bound of power against Standardized Mean Difference (SMD) effects and $112$ observed study sample sizes $N$ in SLR. (\emph{Note:} Only studies with $N < 1250$ are shown for visual clarity, excluding $14$ from the view)}
\label{fig:powerSimCombinedUB}
\end{figure}
}

% DESCRIPTIVES
\newcommand{\descPowerSimCombined}{
\begin{table}[p]
\centering\caption{Descriptives of sample size and upper-bound power distribution}
\label{tab:descPowerSimCombined}
\captionsetup{position=top}
\subfloat[Sample size]{
\label{tab:descSample}
\centering% latex table generated in R 3.6.2 by xtable 1.8-4 package
% Mon Oct  5 16:30:58 2020
\begingroup\footnotesize
\begin{tabular}{rrrrrrrrr}
  \toprule
 & Min & Q:25 & Mdn & Q:75 & Max & IQR & M & SD \\ 
  \midrule
1 & 9.00 & 41.25 & 138.25 & 340.75 & 1627.00 & 299.50 & 264.19 & 339.13 \\ 
   \bottomrule
\end{tabular}
\endgroup
}

\subfloat[Upper bounds of statistical power]{%
\label{tab:descPowerSim}
\centering% latex table generated in R 3.6.2 by xtable 1.8-4 package
% Mon Oct  5 16:30:58 2020
\begingroup\footnotesize
\begin{tabular}{rrrrrrrrrr}
  \toprule
 & smd & Min & Q:25 & Mdn & Q:75 & Max & IQR & M & SD \\ 
  \midrule
1 & 0.20 & 0.06 & 0.10 & 0.21 & 0.45 & 0.98 & 0.36 & 0.31 & 0.26 \\ 
  2 & 0.50 & 0.10 & 0.35 & 0.83 & 1.00 & 1.00 & 0.65 & 0.68 & 0.33 \\ 
  3 & 0.80 & 0.18 & 0.71 & 1.00 & 1.00 & 1.00 & 0.29 & 0.84 & 0.24 \\ 
   \bottomrule
\end{tabular}
\endgroup
}
\end{table}
}

% ES EXTRACTION

\newcommand{\powerSimPlotEST}{
\begin{figure}[tb]

\includegraphics[width=\maxwidth]{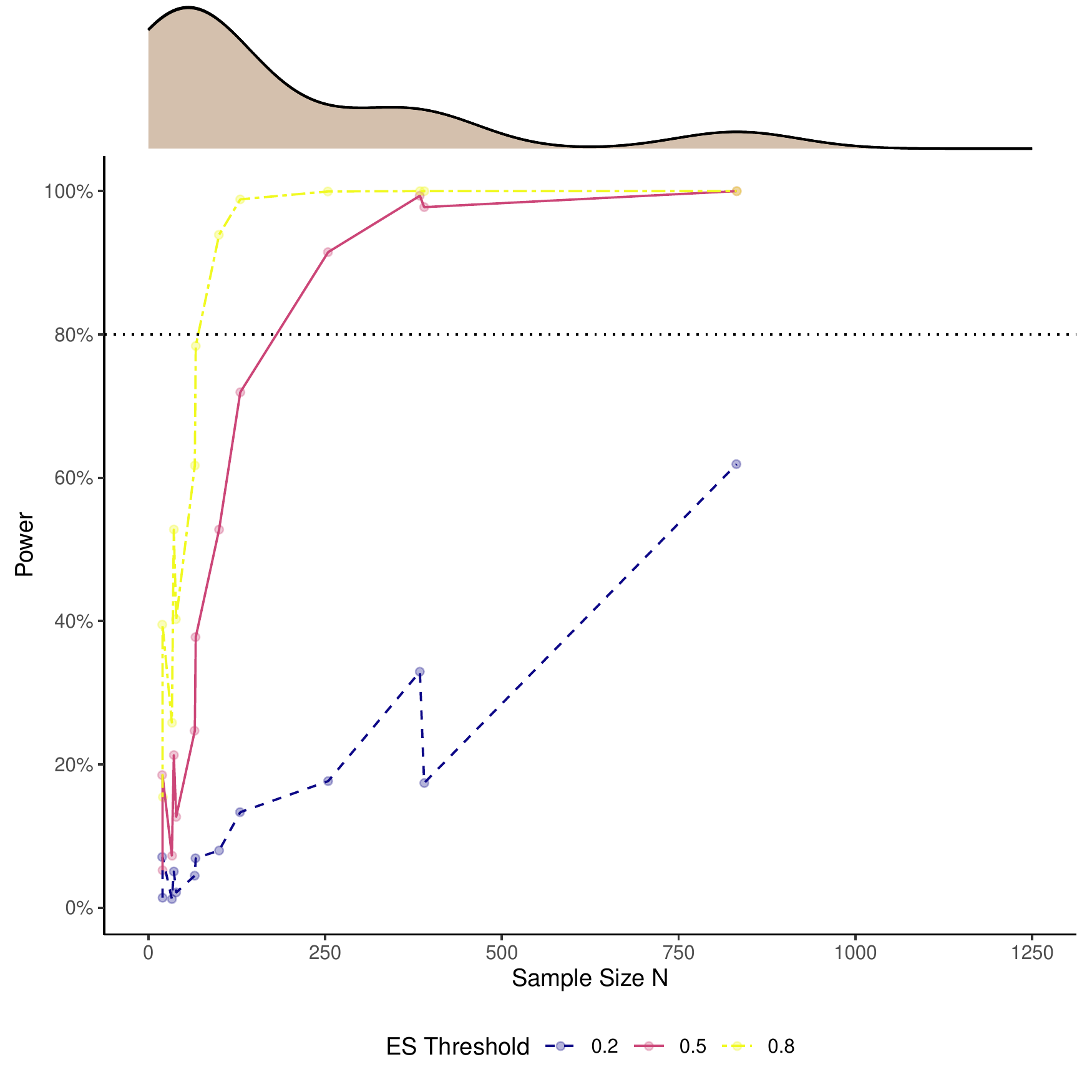} 
\caption{Power simulation for $951$ observed $t$-tests in SLR.}
\label{fig:powerSimPlotEST}
\end{figure}
}

\newcommand{\powerSimCombinedT}{
\begin{figure}[p]
\centering\captionsetup{position=bottom}
\begin{minipage}{0.49\textwidth}%
\subfloat[Power by Sample Size]{
\label{fig:powerSimT}
\centering
\includegraphics[width=\maxwidth]{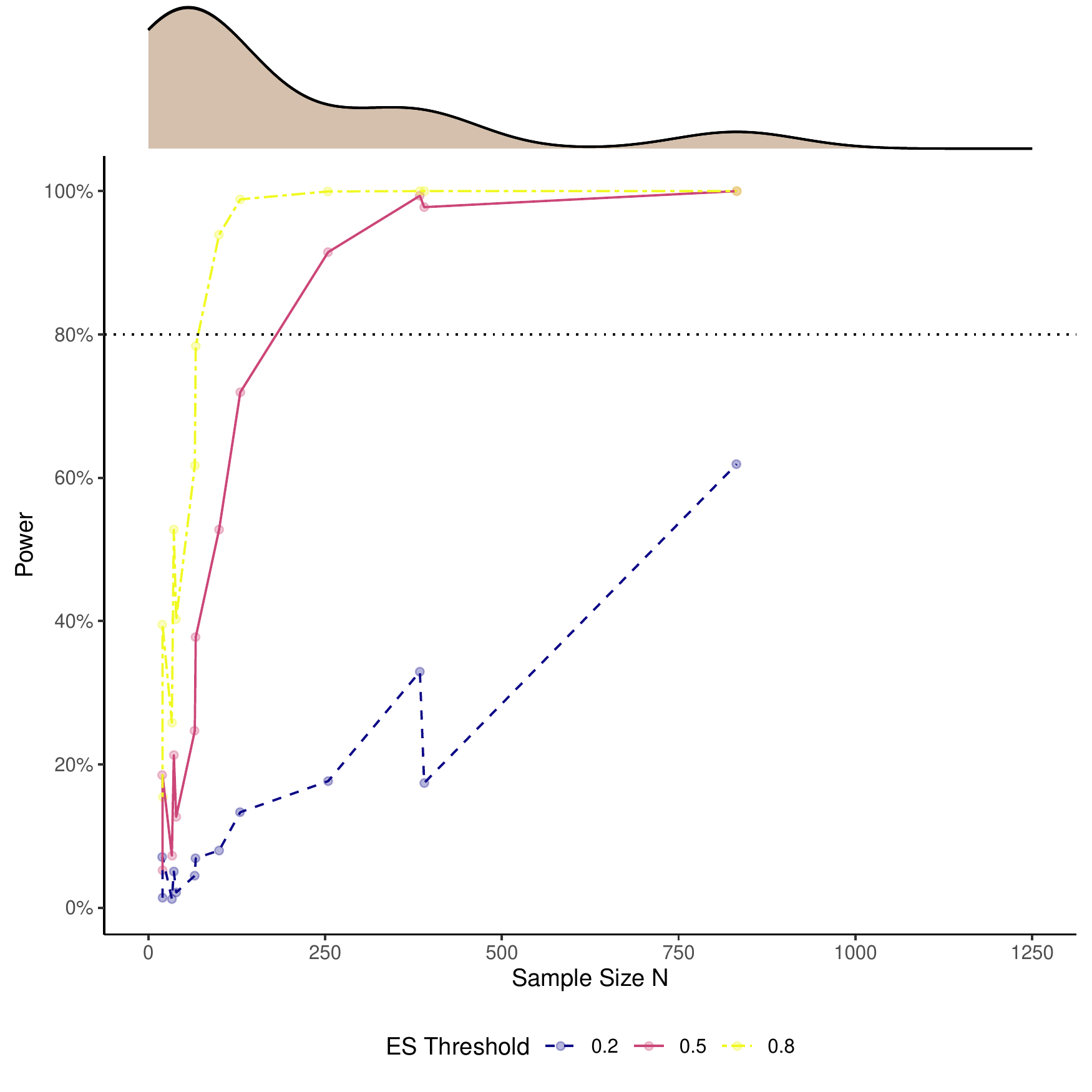} 
}
\end{minipage}~
\begin{minipage}{0.49\textwidth}%
\subfloat[Power Density]{%
\label{fig:powerSimRidgeT}
\centering
\includegraphics[width=\maxwidth]{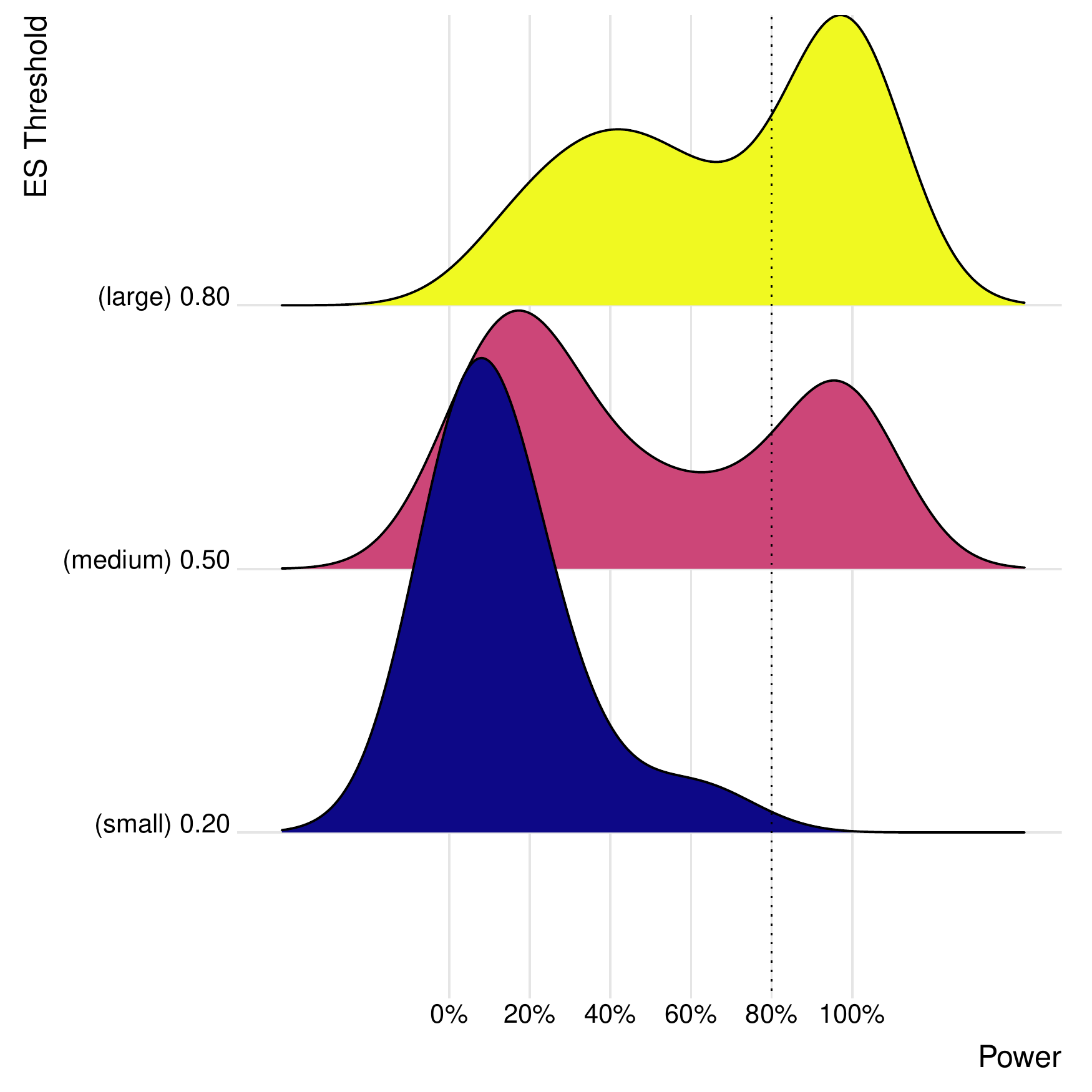} 
}
\end{minipage}
\caption{Power of independent-samples $t$ tests vs. SMD (Hedges' $g$) thresholds}
\label{fig:powerSimCombinedT}
\end{figure}
}

\newcommand{\powerSimCombinedChisq}{
\begin{figure}[p]
\centering\captionsetup{position=bottom}
\begin{minipage}{0.49\textwidth}%
\subfloat[Power by Sample Size]{
\label{fig:powerSimChisq}
\centering
\includegraphics[width=\maxwidth]{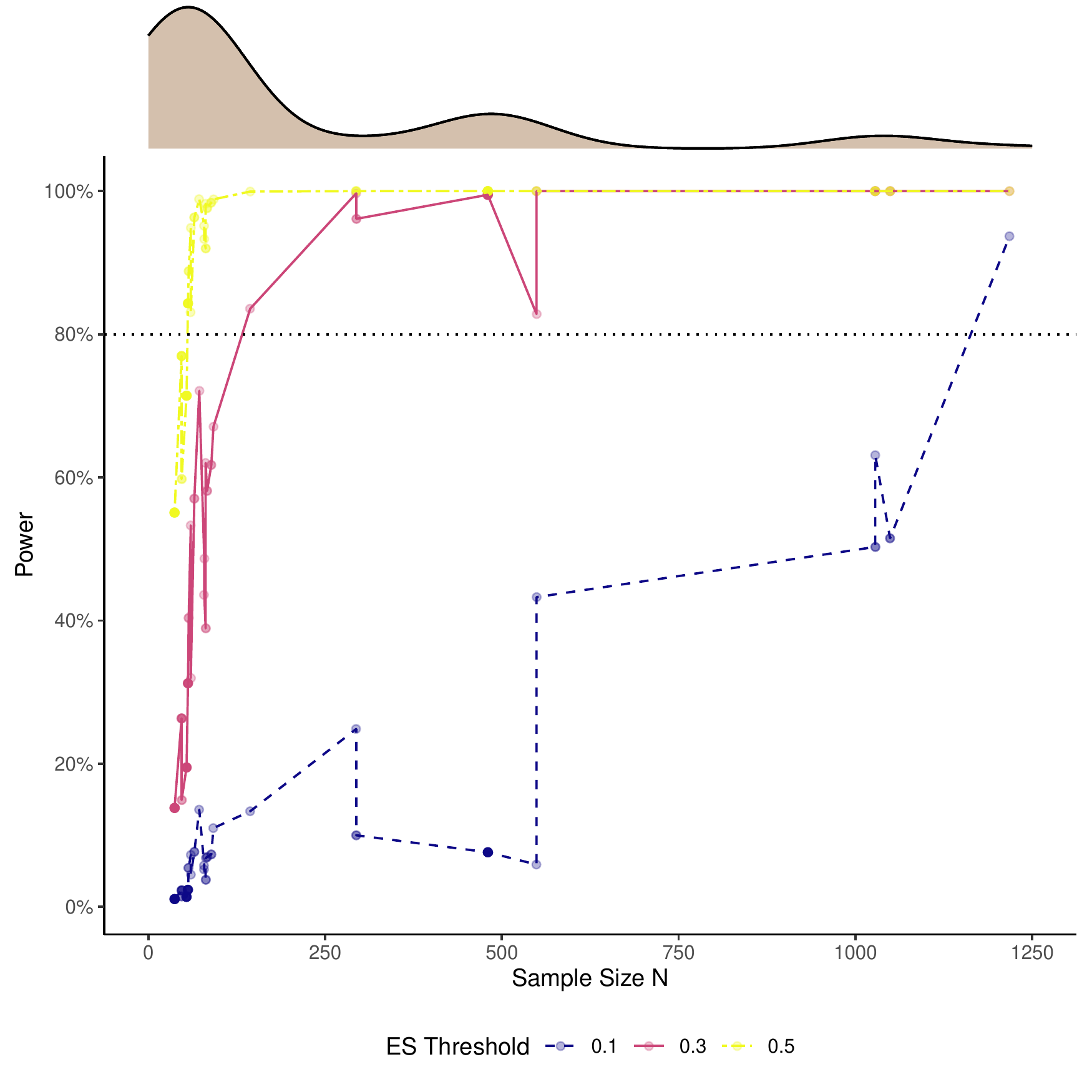} 
}
\end{minipage}~
\begin{minipage}{0.49\textwidth}%
\subfloat[Power Density]{%
\label{fig:powerSimRidgeChisq}
\centering
\includegraphics[width=\maxwidth]{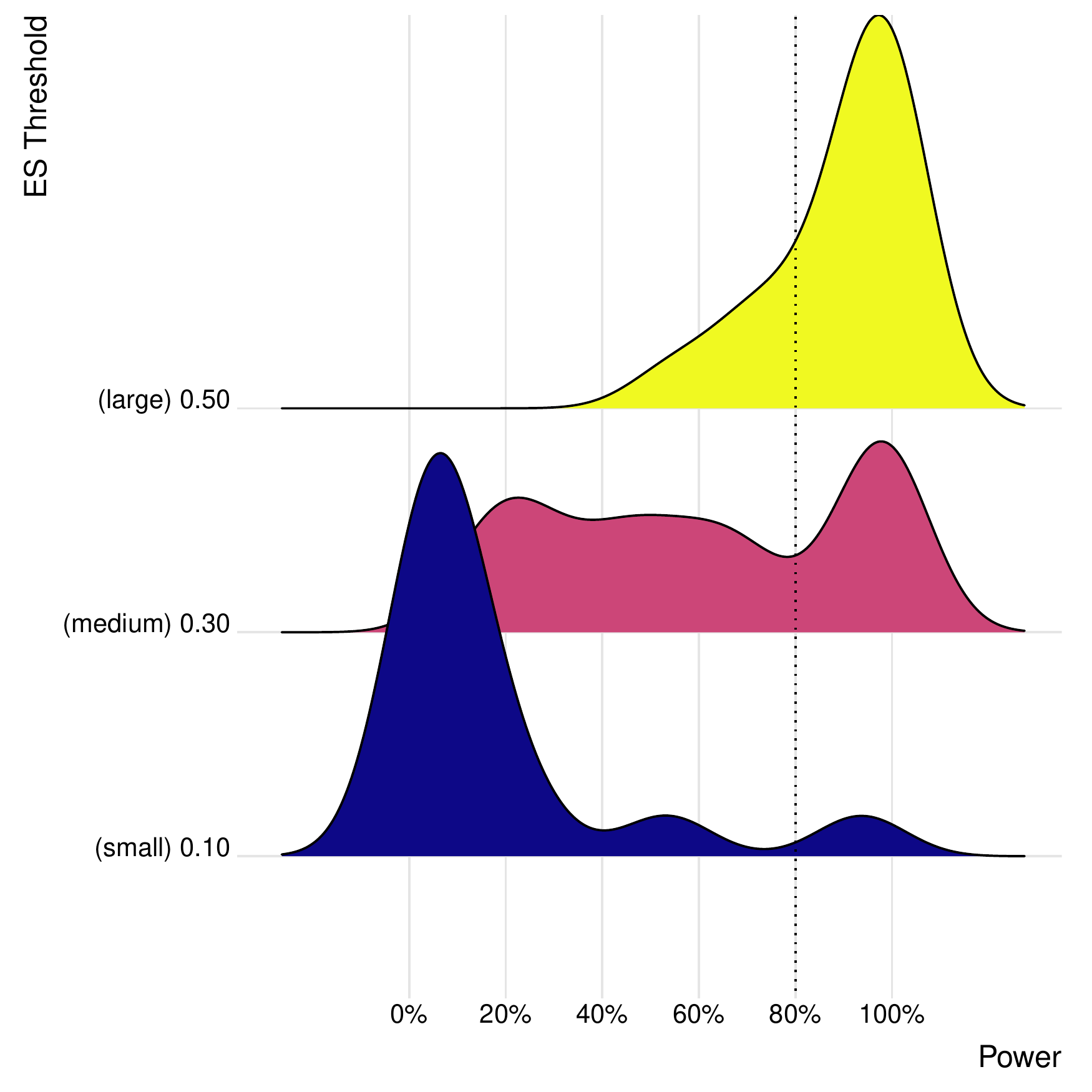} 
}
\end{minipage}
\caption{Power of $\chi^2$ tests vs. Cohen's $w$ thresholds}
\label{fig:powerSimCombinedChisq}
\end{figure}
}

\newcommand{\powerSimCombinedF}{
\begin{figure}[p]
\centering\captionsetup{position=bottom}
\begin{minipage}{0.49\textwidth}%
\subfloat[Power by Sample Size]{
\label{fig:powerSimF}
\centering
\includegraphics[width=\maxwidth]{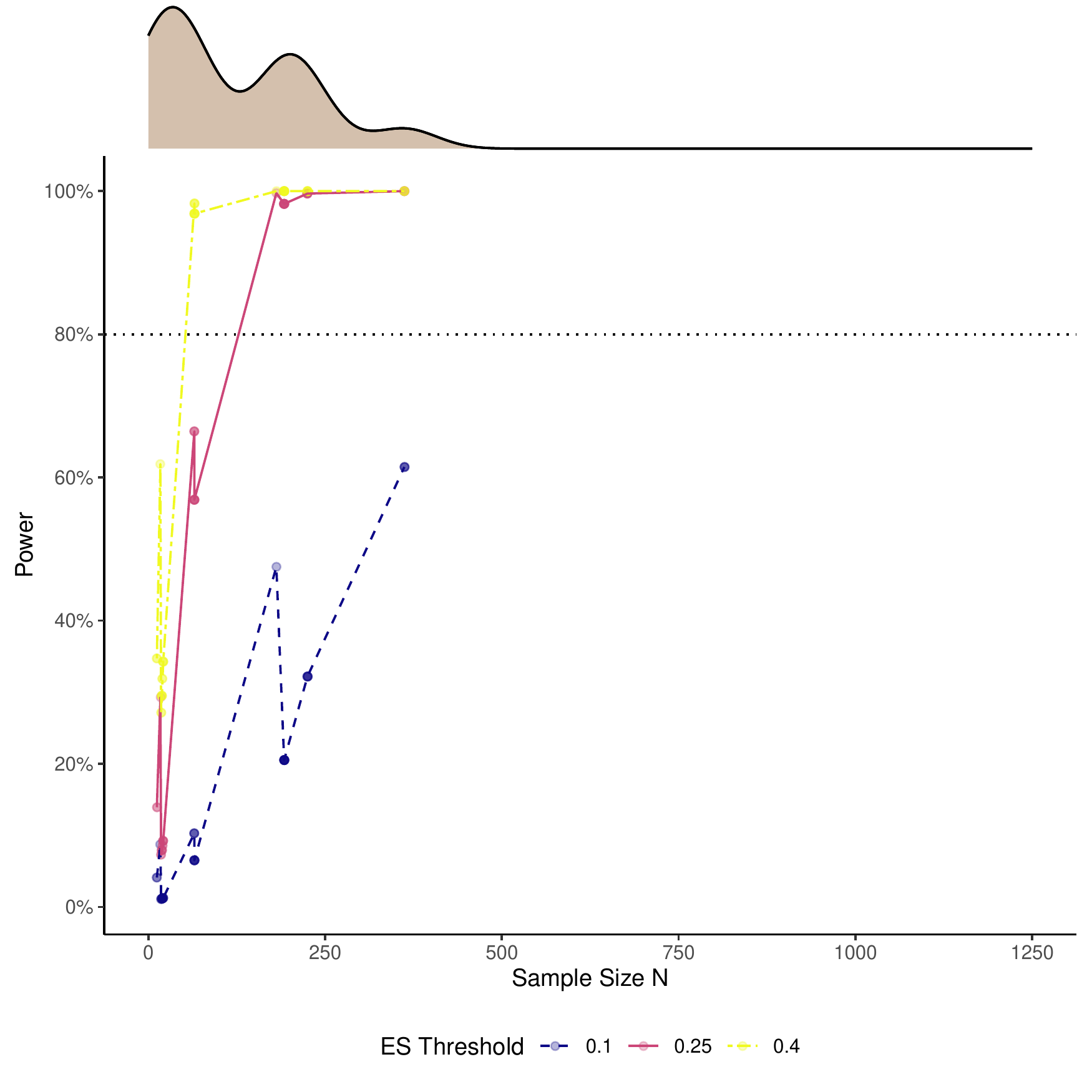} 
}
\end{minipage}~
\begin{minipage}{0.49\textwidth}%
\subfloat[Power Density]{%
\label{fig:powerSimRidgeF}
\centering
\includegraphics[width=\maxwidth]{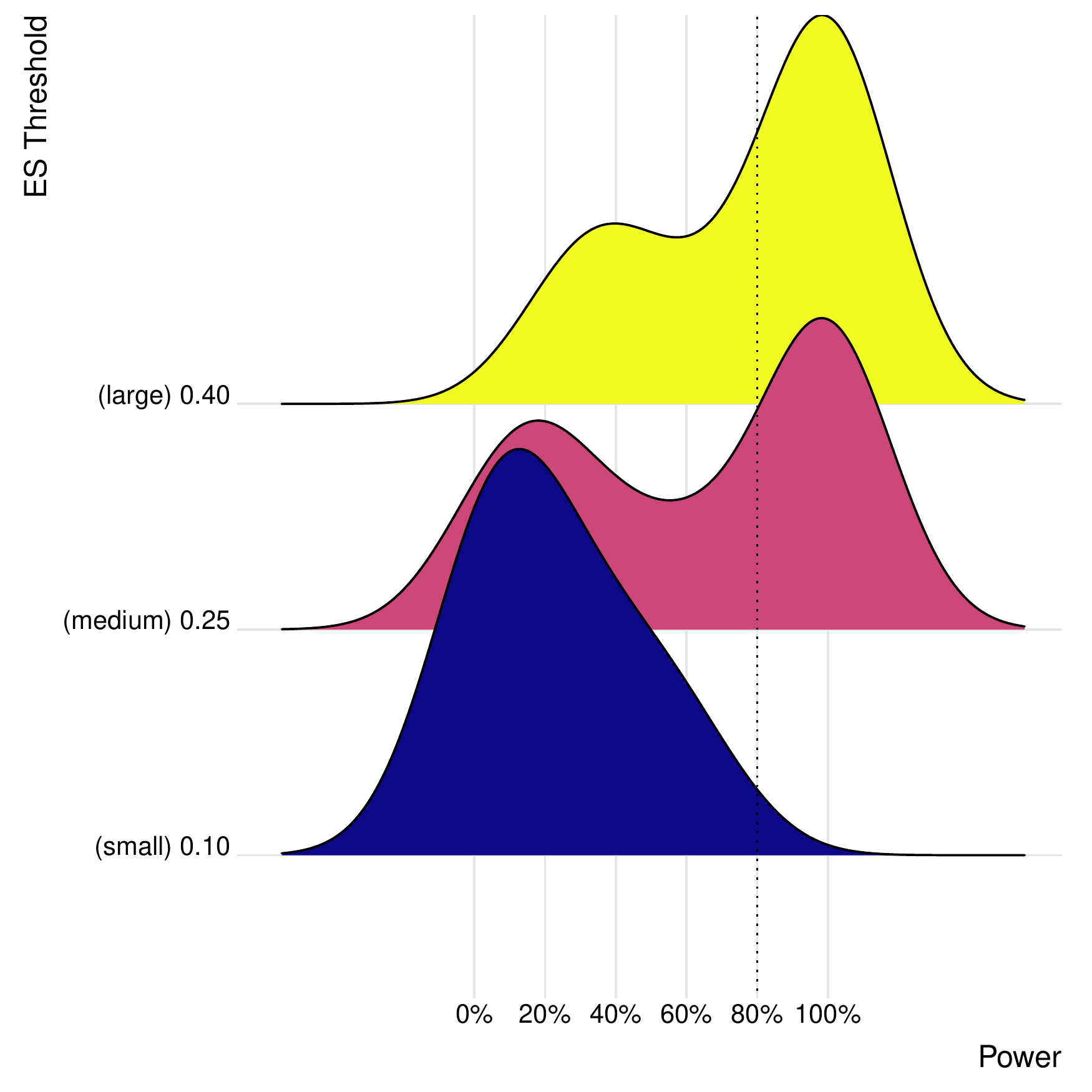} 
}
\end{minipage}
\caption{Power of one-way $F$ tests vs. Cohen's $f$ thresholds}
\label{fig:powerSimCombinedF}
\end{figure}
}

% DESCRIPTIVES
\newcommand{\descExtractedES}{
% latex table generated in R 3.6.2 by xtable 1.8-4 package
% Mon Oct  5 16:31:07 2020
\begin{table}[ht]
\centering
\caption{Descriptives of the observed absolute effect sizes as log odds ratios.} 
\label{tab:descExtractedES}
\begingroup\footnotesize
\begin{tabular}{lrrrrrrrrr}
  \toprule
Statistic & n & Min & Q:25 & Mdn & Q:75 & Max & IQR & M & SD \\ 
  \midrule
Chi2 & 102 & 0.07 & 1.05 & 1.38 & 1.74 & 3.73 & 0.69 & 1.40 & 0.90 \\ 
  F &  42 & 0.04 & 0.31 & 0.73 & 1.57 & 7.71 & 1.26 & 1.43 & 1.81 \\ 
  r &   6 & -1.03 & 1.02 & 1.33 & 1.86 & 2.15 & 0.84 & 1.40 & 0.60 \\ 
  t & 317 & -4.58 & 0.21 & 0.43 & 0.99 & 8.35 & 0.78 & 0.85 & 1.18 \\ 
  Z &  33 & 0.07 & 0.31 & 0.66 & 1.59 & 7.17 & 1.29 & 1.37 & 1.57 \\ 
   \bottomrule
\end{tabular}
\endgroup
\end{table}
}
\newcommand{\descPowerSimESCombined}{
\begin{table}[p]
\centering\caption{Descriptives of statistical power of actual tests with and without MCC}
\label{tab:descPowerSimESCombined}
\captionsetup{position=top}
\subfloat[Without MMC]{
\label{tab:descPowerSimES}
\centering% latex table generated in R 3.6.2 by xtable 1.8-4 package
% Mon Oct  5 16:31:07 2020
\begingroup\footnotesize
\begin{tabular}{rlrrrrrrrr}
  \toprule
 & threshold & Min & Q:25 & Mdn & Q:75 & Max & IQR & M & SD \\ 
  \midrule
1 & large & 0.40 & 0.80 & 0.98 & 1.00 & 1.00 & 0.20 & 0.86 & 0.20 \\ 
  2 & medium & 0.19 & 0.41 & 0.68 & 1.00 & 1.00 & 0.59 & 0.66 & 0.31 \\ 
  3 & small & 0.07 & 0.10 & 0.12 & 0.53 & 1.00 & 0.44 & 0.34 & 0.34 \\ 
   \bottomrule
\end{tabular}
\endgroup
}

\subfloat[With MCC]{%
\label{tab:descPowerSimESMCC}
\centering% latex table generated in R 3.6.2 by xtable 1.8-4 package
% Mon Oct  5 16:31:07 2020
\begingroup\footnotesize
\begin{tabular}{rlrrrrrrrr}
  \toprule
 & threshold & Min & Q:25 & Mdn & Q:75 & Max & IQR & M & SD \\ 
  \midrule
1 & large & 0.15 & 0.62 & 0.94 & 1.00 & 1.00 & 0.38 & 0.79 & 0.27 \\ 
  2 & medium & 0.05 & 0.22 & 0.53 & 1.00 & 1.00 & 0.77 & 0.57 & 0.37 \\ 
  3 & small & 0.01 & 0.04 & 0.08 & 0.28 & 1.00 & 0.24 & 0.24 & 0.33 \\ 
   \bottomrule
\end{tabular}
\endgroup
}
\end{table}
}

\newcommand{\caterpillarES}{
\begin{figure}[tbp]
\vspace{-1cm}

\includegraphics[width=\maxwidth]{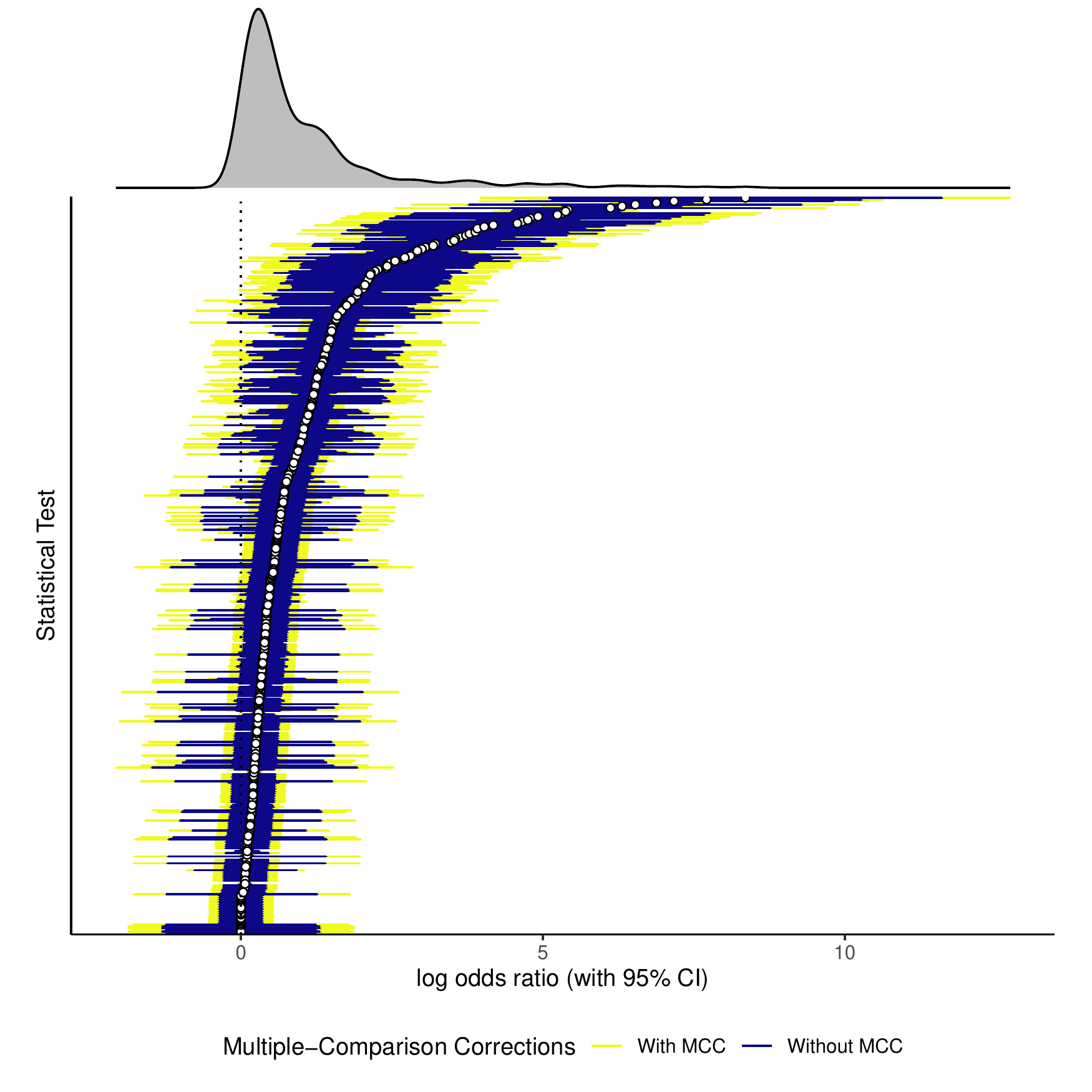} 
\vspace{-1cm}
\caption{Caterpillar forest plot of $n=431$ log odds ratios and their 95\% confidence intervals, ordered by log(OR). \processifversion{DocumentVersionTR}{\emph{Note}: The dark-blue error bar represents the unmodified 95\% CI; the light-yellow error bar represents the Bonferroni-corrected 95\% CI.}}
\label{fig:caterpillarES}
\end{figure}
}
\newcommand{\descSigMCCcombined}{
\begin{table}[p]
\centering\caption{Absolute and before-after matched contingency tables of significance by multiple-comparison corrections (MCC)}
\captionsetup{position=top}
\label{tab:descSigMCCcombined}
\subfloat[Overall Frequencies]{
\label{tab:descSigMCC}
\centering% latex table generated in R 3.6.2 by xtable 1.8-4 package
% Mon Oct  5 16:31:10 2020
\begingroup\footnotesize
\begin{tabular}{rrr}
  \toprule
 & Significant & Not Significant \\ 
  \midrule
MCC & 165 & 266 \\ 
  No MCC & 232 & 199 \\ 
   \bottomrule
\end{tabular}
\endgroup
}~\subfloat[Matched-Pairs Before/After MCC]{%
\label{tab:descSigMCCmatched}
\centering% latex table generated in R 3.6.2 by xtable 1.8-4 package
% Mon Oct  5 16:31:10 2020
\begingroup\footnotesize
\begin{tabular}{rrr}
  \toprule
 & Significant & Not Significant \\ 
  \midrule
Significant & 165 &  67 \\ 
  Not Significant &   0 & 199 \\ 
   \bottomrule
\end{tabular}
\endgroup
}
\end{table}
}

\newcommand{\powerDensityMCC}{
\begin{figure}[tbp]
\vspace{-2.5cm}

\includegraphics[width=\maxwidth]{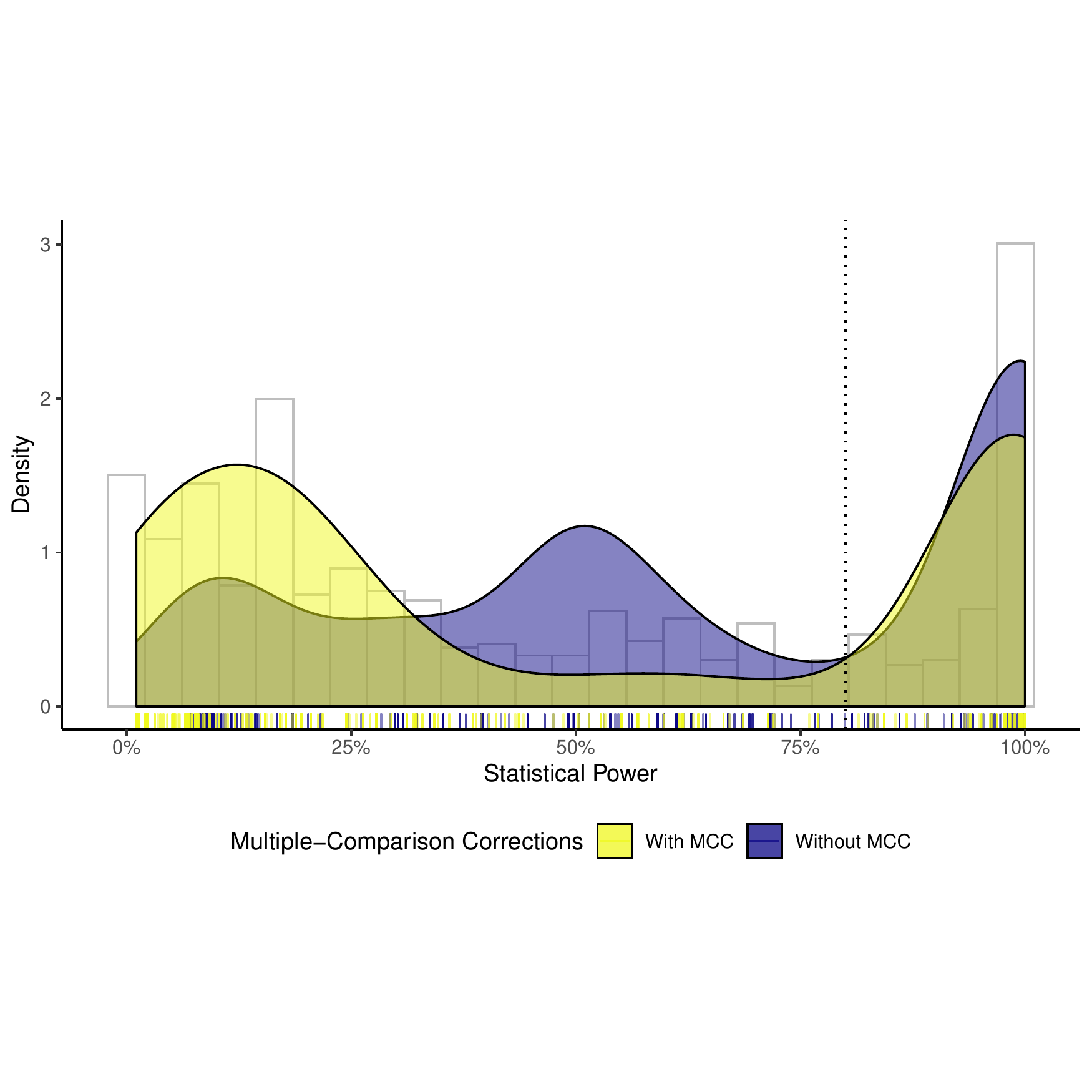} 
\vspace{-2.5cm}
\caption{Histogram-density plot comparing statistical power for all tests in the sample by MCC. 
           \emph{Note:} The histogram is with MCC and square-root transformed.}
\label{fig:powerDensityMCC}
\end{figure}
}
\newcommand{\powerDensityMCCthresholds}{
\begin{figure*}[tbp]
\centering\captionsetup{position=bottom}
\subfloat[small]{
\label{fig:powerDensityMCCthresholdsSmall}
\centering\begin{minipage}{.8\textwidth}%
\vspace{-2cm}

\includegraphics[width=\maxwidth]{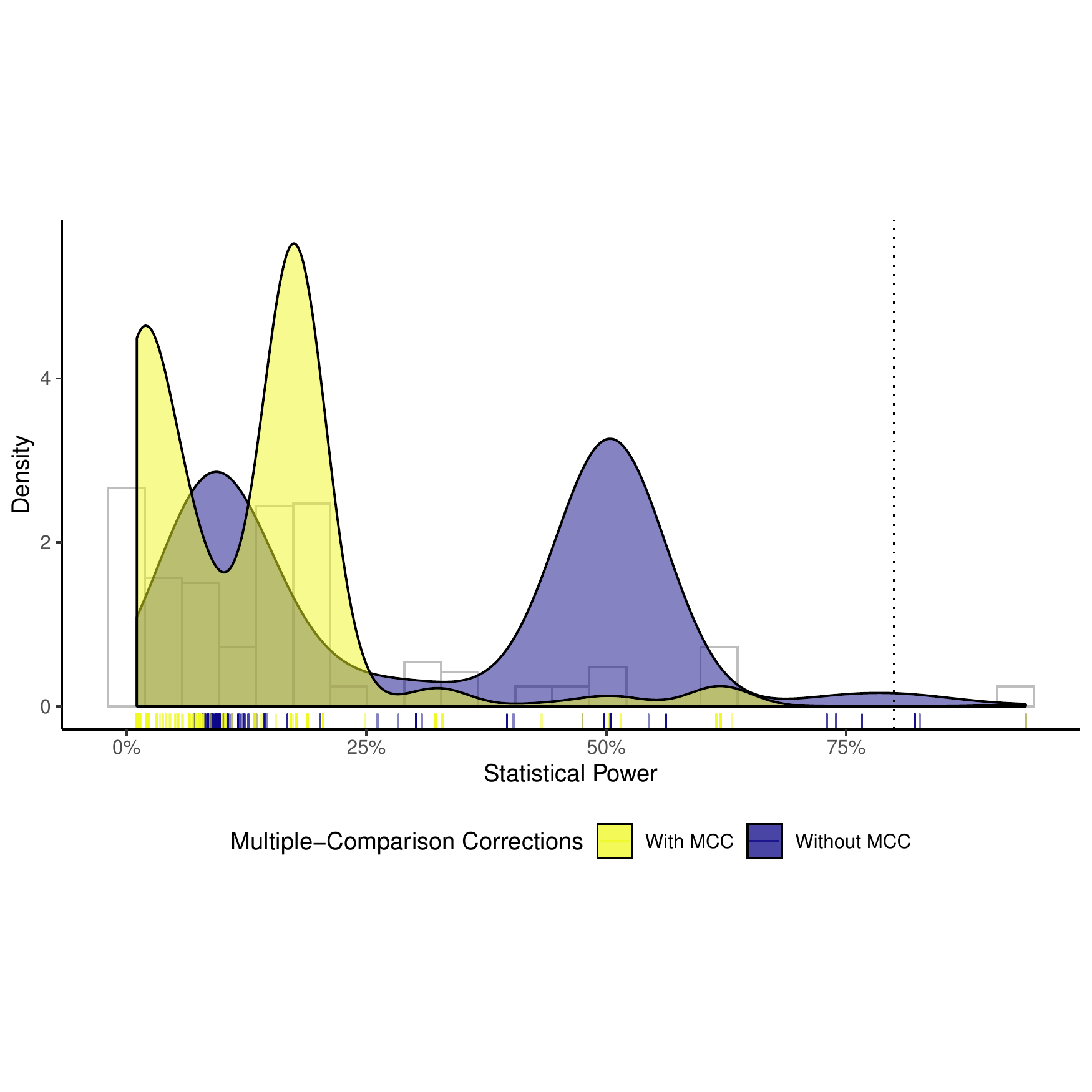} 
\vspace{-2.5cm}
\end{minipage}
}

\subfloat[medium]{%
\label{fig:powerDensityMCCthresholdsMedium}
\centering\begin{minipage}{.8\textwidth}%
\vspace{-2cm}

\includegraphics[width=\maxwidth]{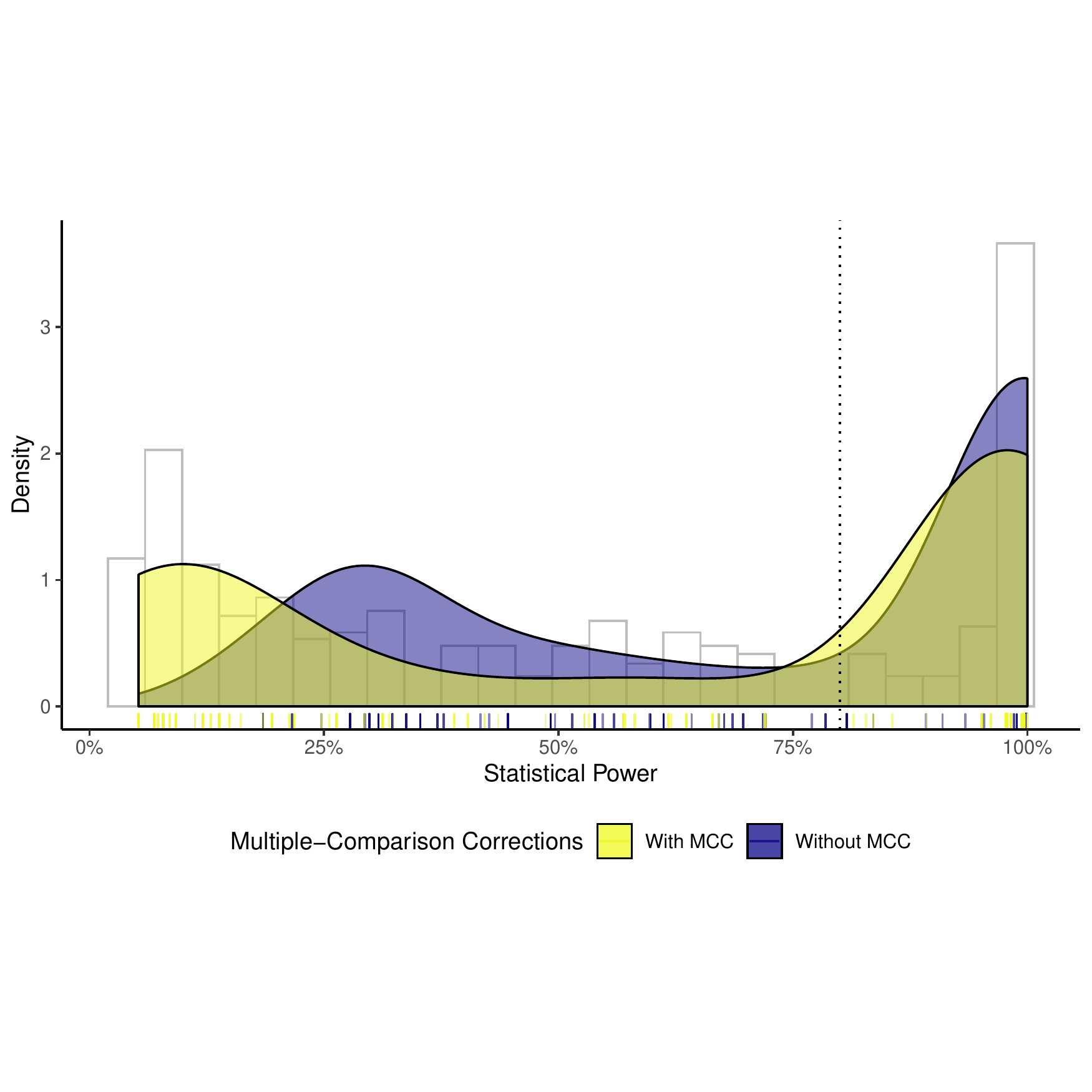} 
\vspace{-2.5cm}
\end{minipage}
}

\subfloat[large]{%
\label{fig:powerDensityMCCthresholdsLarge}
\centering\begin{minipage}{.8\textwidth}%
\vspace{-2cm}

\includegraphics[width=\maxwidth]{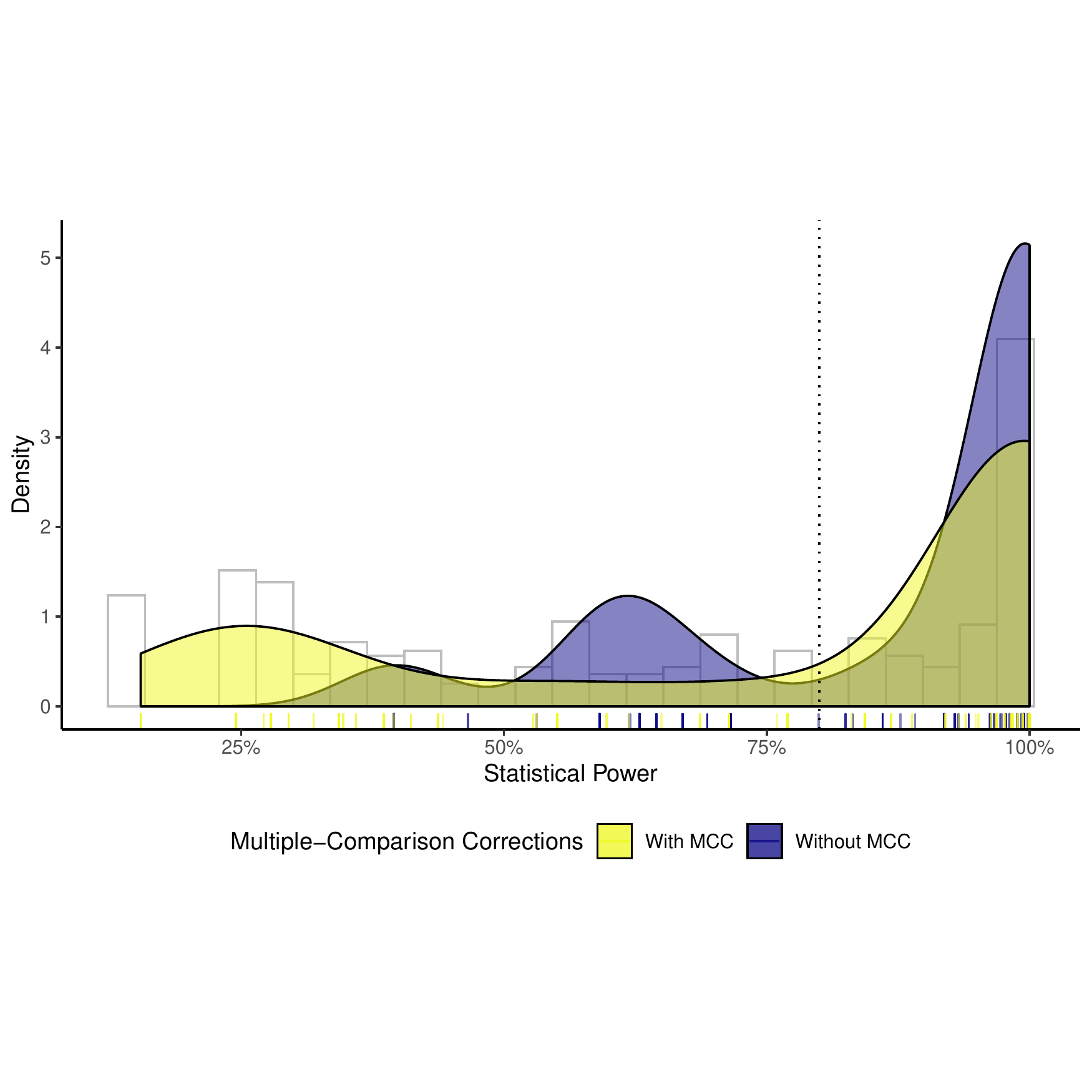} 
\vspace{-2.5cm}
\end{minipage}
}
\caption{Histogram-density plot comparing statistical power by effect size thresholds without MMC and with MCC. 
           \emph{Note:} The histogram square-root transformed for visual clarity.}
\label{fig:powerDensityMCCthresholds}
\end{figure*}
}

\newcommand{\winnersCurse}{
\begin{figure}[tb]
\vspace{-2.5cm}

\includegraphics[width=\maxwidth]{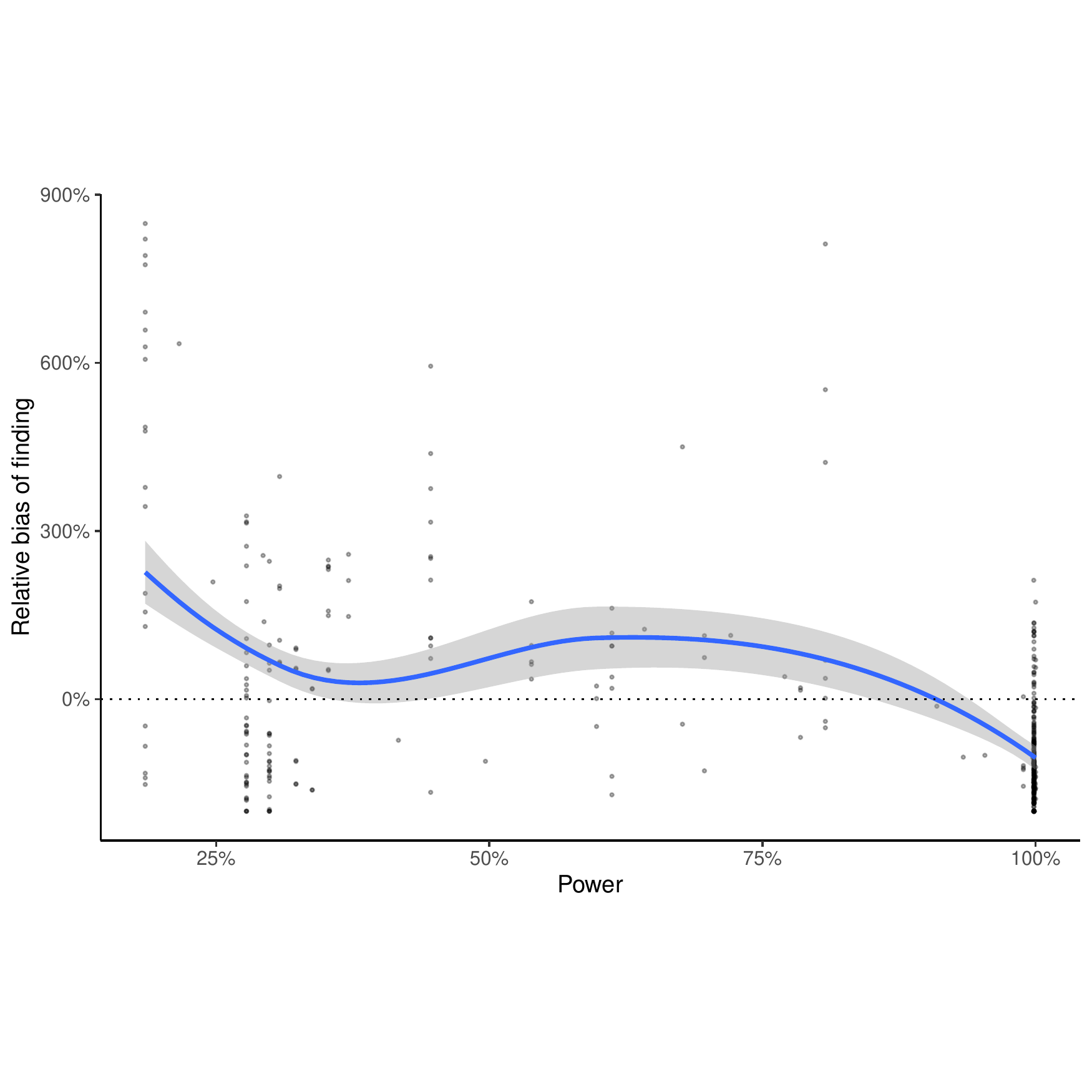} 
\vspace{-2.5cm}
\caption{ES Bias by Power illustrating the Winner's Curse. \emph{Note:} Entries with more than 1000\% bias were removed for visual clarity without impact on the result.}
\label{fig:winnersCurse}
\end{figure}
}
\newcommand{\winnersCurseResidual}{
\begin{figure}[tb]

\includegraphics[width=\maxwidth]{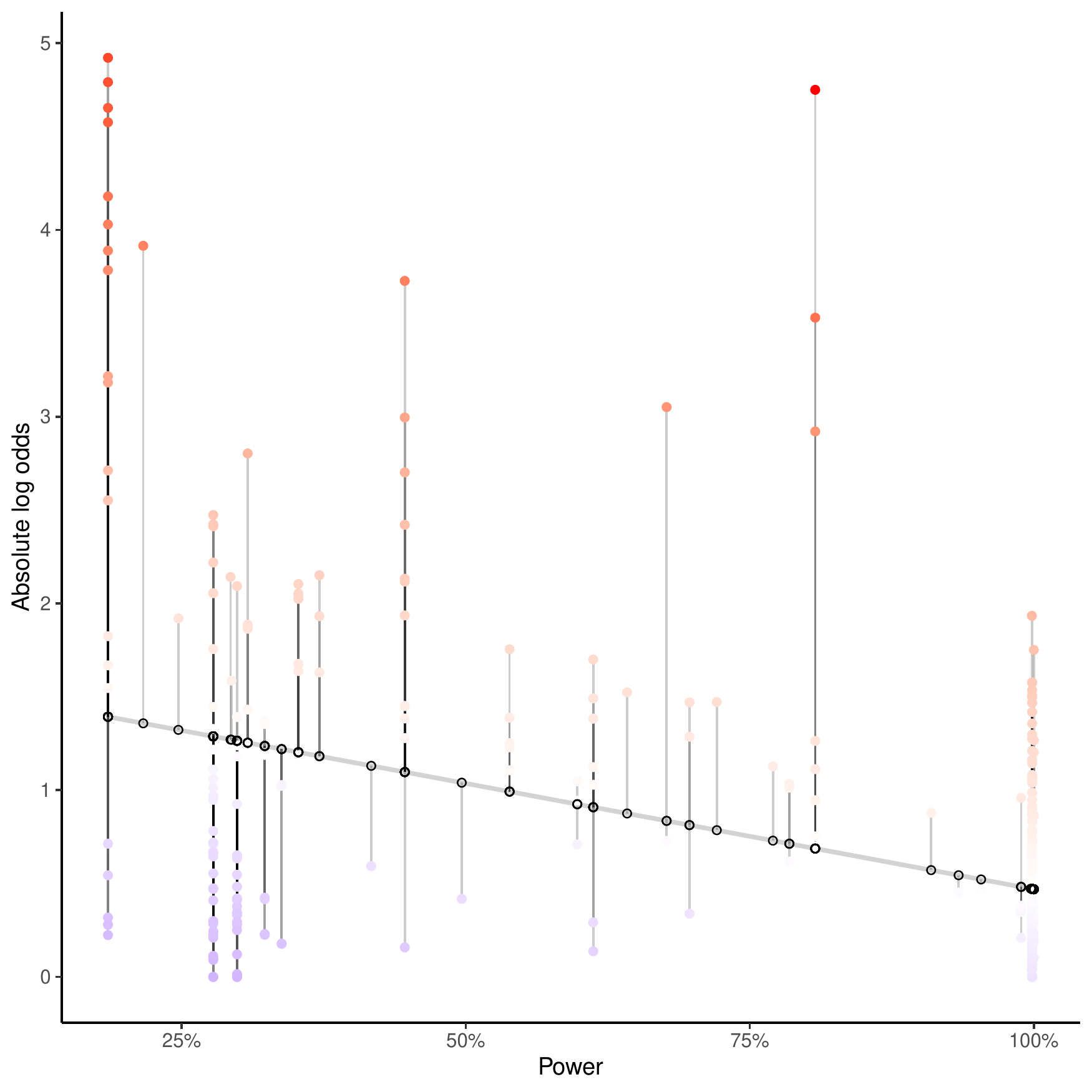} 
\caption{Robust regression of absolute log odds by power and its residuals}
\label{fig:winnersCurseResidual}
\end{figure}
}

\newcommand{\combinedFunnelPlots}{
\begin{figure}[tbp]
\centering\captionsetup{position=bottom}
\begin{minipage}{0.49\textwidth}%
\subfloat[Per paper (Mean ES and SE)]{
\label{fig:funnelPlotAvg}
\centering
\includegraphics[width=\maxwidth]{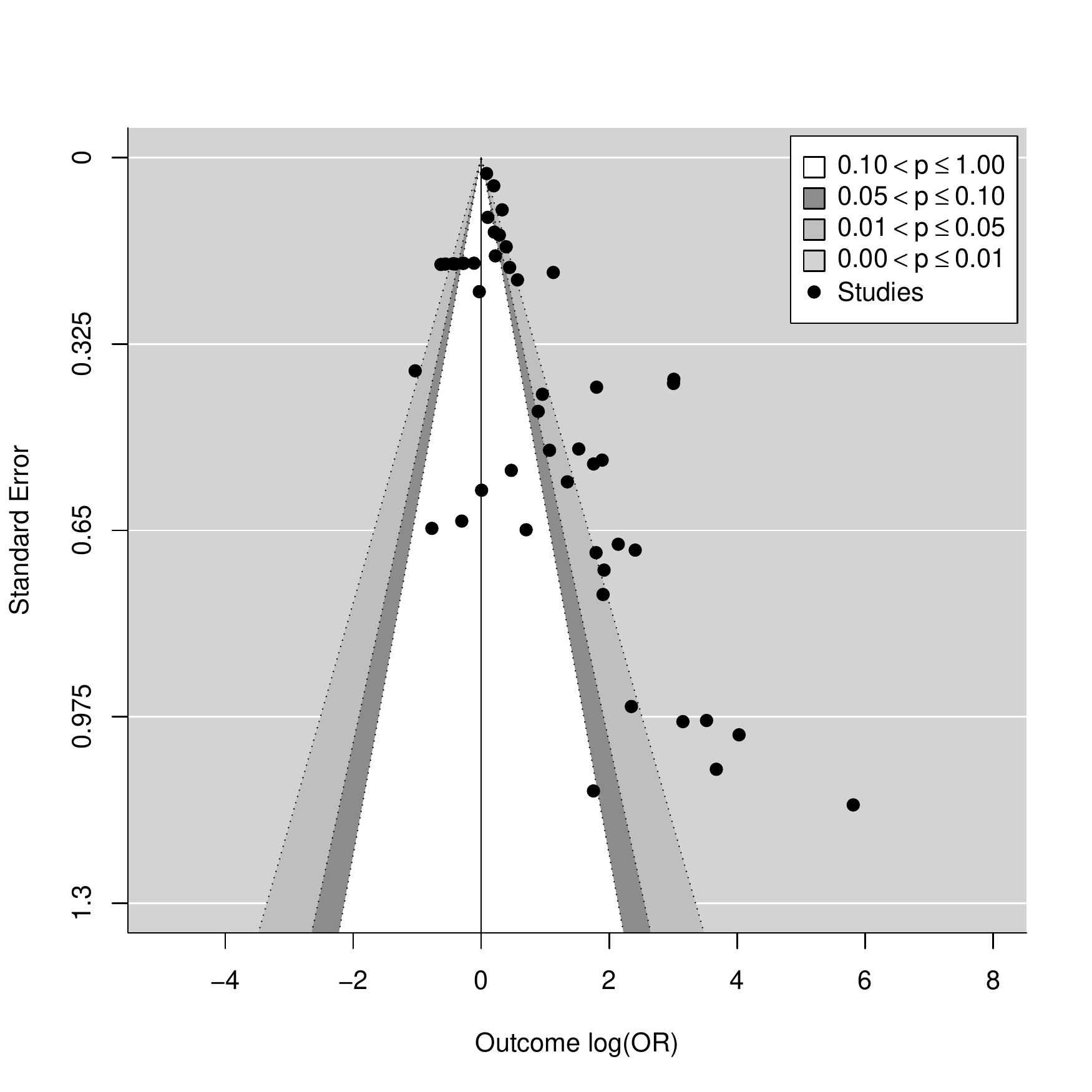} 
}
\end{minipage}~
\begin{minipage}{0.49\textwidth}%
\subfloat[Per test]{%
\label{fig:funnelPlotAll}
\centering
\includegraphics[width=\maxwidth]{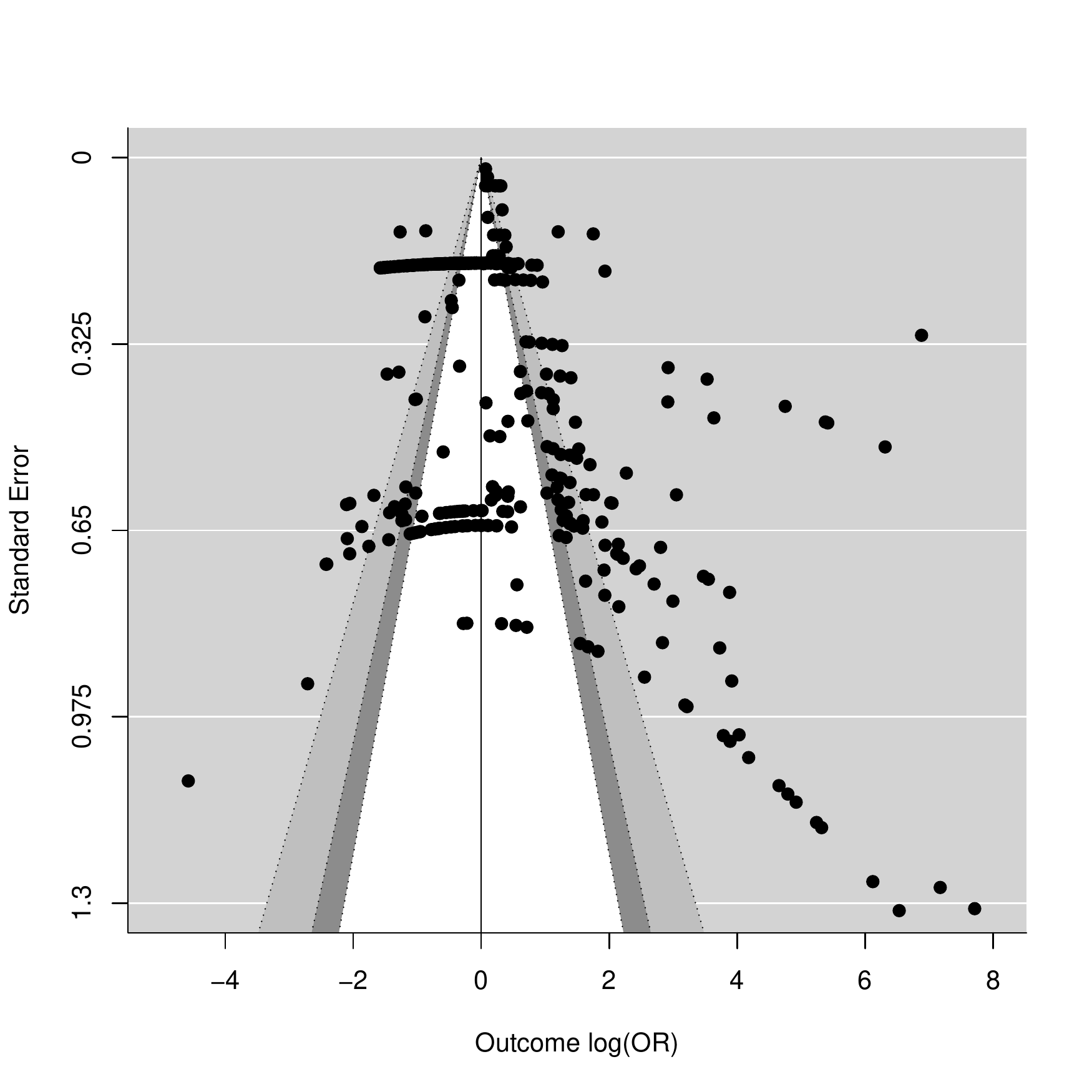} 
}
\end{minipage}
\caption{Funnel plots of \textsf{log}(\textit{OR}) effect sizes and their standard errors}
\label{fig:combinedFunnelPlots}
\end{figure}
}

\newcommand{\sigChasing}{
\begin{figure*}[tbp]
\centering\captionsetup{position=bottom}
\begin{minipage}{0.49\textwidth}%
\subfloat[Uncorrected]{
\label{fig:sigChasing.NoMCC}
\centering
\includegraphics[width=\maxwidth]{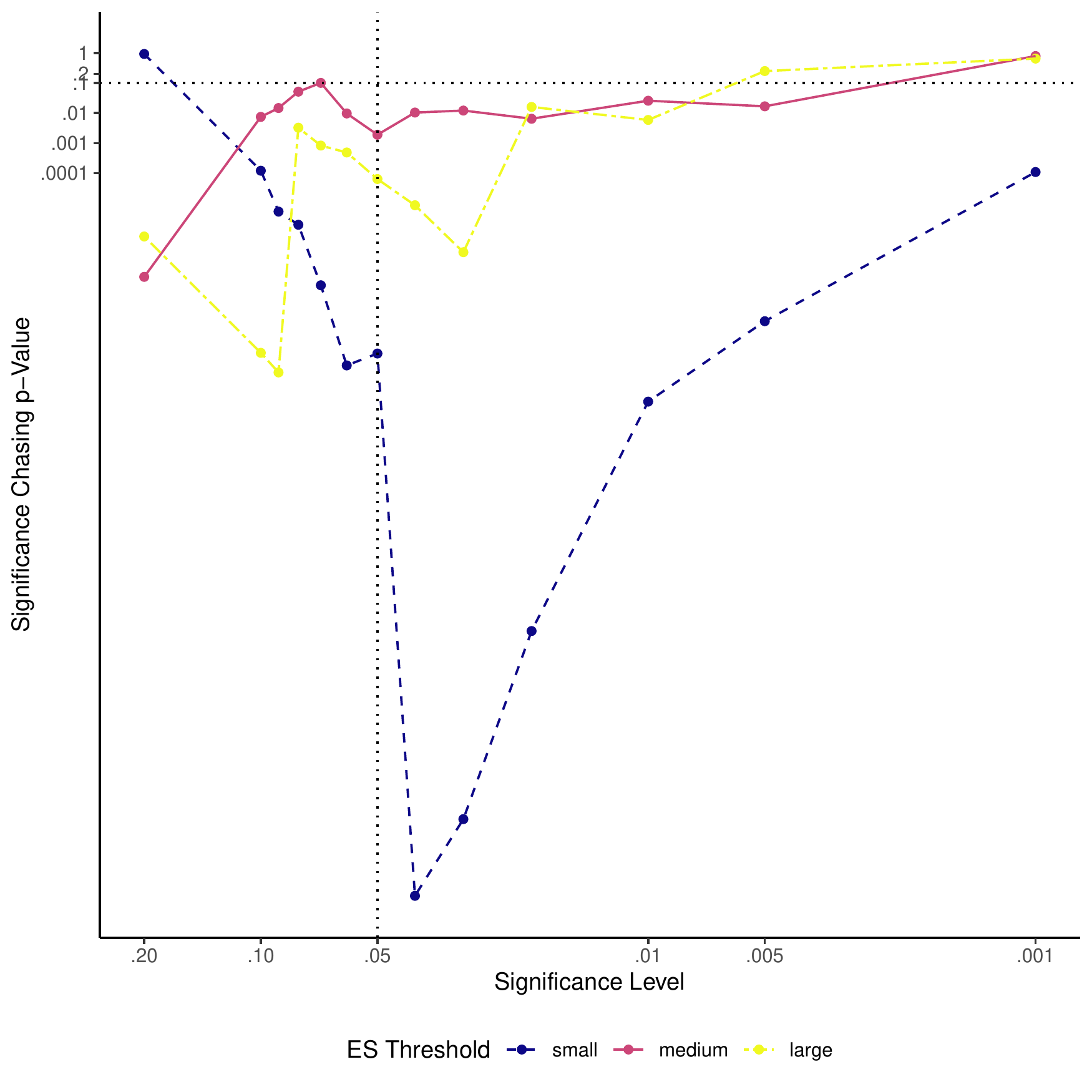} 
}
\end{minipage}~
\begin{minipage}{0.49\textwidth}%
\subfloat[With MCC]{%
\label{fig:sigChasing.MCC}
\centering
\includegraphics[width=\maxwidth]{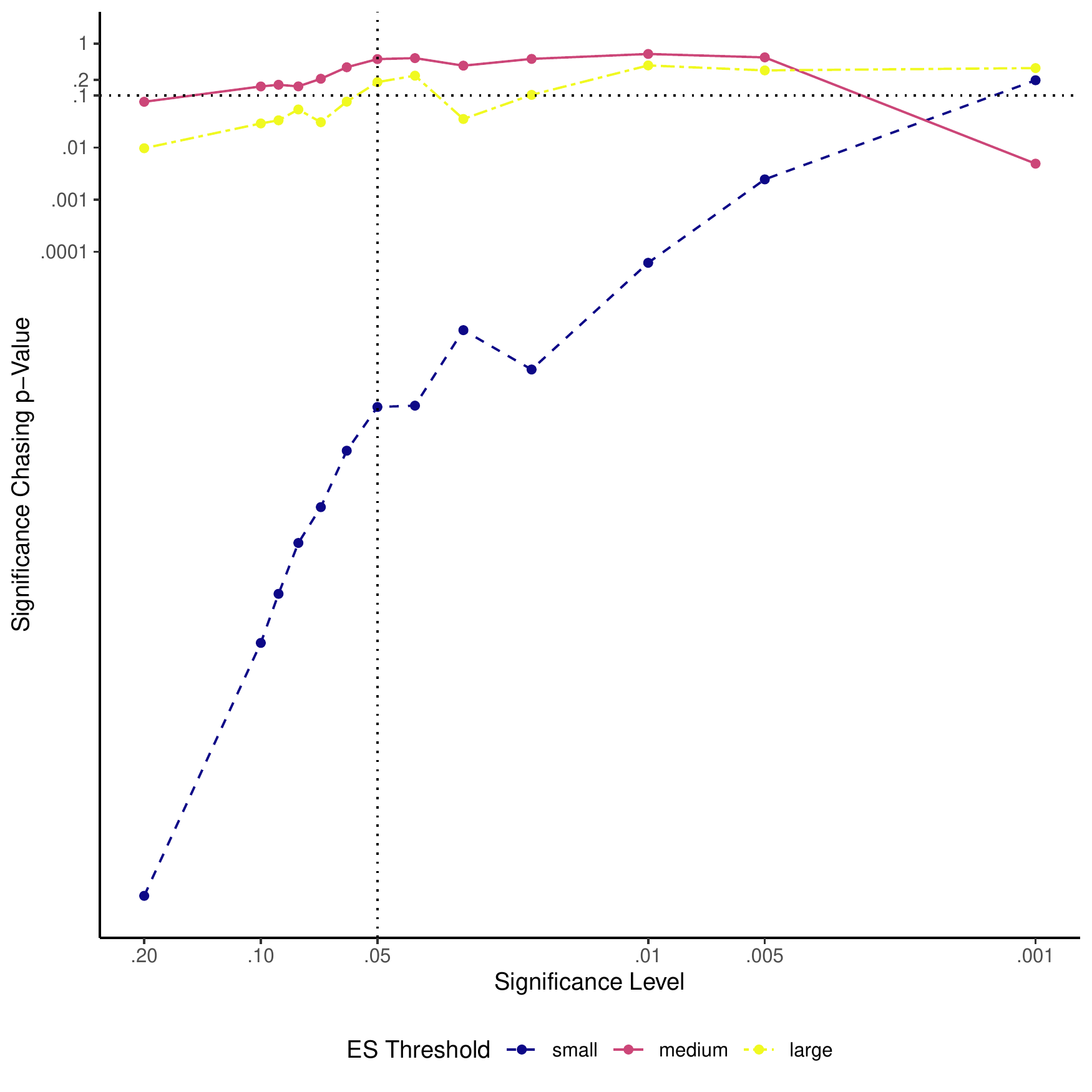} 
}
\end{minipage}
\caption{Significance chasing $\chi^2$ $p$-value by significance level $\alpha$, with and without MCC}
\label{fig:sigChasing}
\end{figure*}
}

% ----------------------------------------

%don't want date printed
\date{}

\begin{DocumentVersionConference}
\title{Statistical Reliability of 10 Years of Cyber Security User Studies\thanks{Open Science Framework: \protect\url{osf.io/bcyte}}}
\end{DocumentVersionConference}
\begin{DocumentVersionTR}
\title{Statistical Reliability of 10 Years of Cyber Security User Studies\thanks{Open Science Framework: \protect\url{osf.io/bcyte}. The definitive short version of this paper is to appear in Proceedings of the 10\textsuperscript{th} International Workshop on Socio-Technical Aspects in Security (STAST 2020), LNCS 11739, Springer, 2020.}\\{}
[Extended Version]}
\end{DocumentVersionTR}

\author{Thomas Gro{\ss}}
\institute{School of Computing\\Newcastle University, UK\\
\email{thomas.gross@newcastle.ac.uk}}
%\authornote{}
%\orcid{1234-5678-9012}

\maketitle

\begin{abstract}
\noindent\textbf{Background.}
  % State the background and context of the work described in the paper.
In recent years, cyber security security user studies have been appraised in meta-research, mostly focusing on the completeness of their statistical inferences and the fidelity of their statistical reporting. However, estimates of the field's distribution of statistical power and its publication bias have not received much attention.

\noindent\textbf{Aim.}
  % State the research question, objective, or purpose of the work in the paper.
In this study, we aim to estimate the effect sizes and their standard errors present as well as the implications on statistical power and publication bias.

\noindent\textbf{Method.}
  % Briefly summarize the method used to conduct the research, including the subjects, procedure, data, and analytical method.
We built upon a published systematic literature review of $146$ user studies in cyber security (2006--2016). We took into account $431$ statistical inferences including $t$-, $\chi^2$-, $r$-, one-way $F$-tests, and $Z$-tests. In addition, we coded the corresponding total sample sizes, group sizes and test families. Given these data, we established the observed effect sizes and evaluated the overall publication bias. We further computed the statistical power vis-{\`a}-vis of parametrized population thresholds to gain unbiased estimates of the power distribution.

\noindent\textbf{Results.}
  % State the outcome of the research using measures appropriate for the study conducted. Results are essentially the numbers.
We obtained a distribution of effect sizes and their conversion into comparable log odds ratios together with their standard errors. We, further, gained funnel-plot estimates of the publication bias present in the sample as well as insights into the power distribution and its consequences.

\noindent\textbf{Conclusions.}
  %  State the anticipated impact. The conclusions are the ``so what'' of the study.
  %  Impact
Through the lenses of power and publication bias, we shed light on the statistical reliability of the studies in the field.
The upshot of this introspection is practical recommendations on conducting and evaluating studies to advance the field.

\keywords{User studies \and SLR \and Cyber security \and Effect Estimation \and Statistical Power \and Publication Bias \and Winner's Curse}
\end{abstract}

% Registration OSF: osf.io/bcyte/

\section{Introduction}
Cyber security user studies and quantitative studies in socio-technical aspects of security in general often rely on statistical inferences to make their case that observed effects are not down to chance. They are to separate the wheat from the chaff.
Indeed, null hypothesis significance testing and $p$-values indicating statistical significance hold great sway in the community.
While the studies in the field have been appraised in recent years on the completeness and fidelity of their statistical reporting, we may still ask how reliable the underlying statistical inferences really are.

``\emph{To what extent can we rely on reported effects?}''
This question can take multiple shapes.
First, we may consider the magnitude of observed effects. While a statement of statistical significance is dependent on the sample size at which the inference was obtained, the magnitude of an effect, its \emph{effect size}, informs us whether an effect is practically relevant---or not.
While small effects might not make much difference in practice and might not be economical to pursue, large effects estimated with confidence can guide us to the interventions that are likely carrying considerable weight in socio-technical systems.

Indeed, a second dimension of reliability pertains to the confidence we have in observed effects, typically measured with 95\% confidence intervals. Here, we are interested how tightly the confidence interval envelops the effect point estimate. The rationale behind such a confidence interval is that if an experiment were repeated many times, we would expect 95\% of the observed effect estimates to be within the stated confidence intervals. Wide intervals, thereby, give us little confidence in the accuracy of an estimation procedure. 

This consideration is exacerbated if a study conducted many tests in the same test family. Given the risk of multiple comparisons to amplify false-positive rates, we would need to adjust the confidence intervals accounting for the multiplicity and, hence, be prepared to gain even less confidence in the findings.

Third, we may consider statistical power, the likelihood of finding an effect that is present in reality. To put it in precise terms, it is the likelihood of rejecting a null hypothesis when it is, in fact, false---the complement of the false negative rate. At the same time, statistical power also impacts the likelihood that a positive report is actually true, hence further impacts the reliability of a finding.
The power distribution, further, offers a first assessment on the statistical reliability of the field.

Finally, we expand on the reliability of the field in terms of evaluating research biases that could undermine results.
\begin{DocumentVersionConference}
Two predominant biases of interest are 
\begin{inparaenum}[(i)]
\item the publication bias~\cite{rosenthal1979file}, and
\item the related winner's curse~\cite{button2013power}.
\end{inparaenum}
\end{DocumentVersionConference}
\begin{DocumentVersionTR}
Three biases of interest are 
\begin{inparaenum}[(i)]
\item the publication bias~\cite{rosenthal1979file,light1984summing,dickersin1990existence,sterling1995publication},
\item the related winner's curse~\cite{button2013power}, and
\item significance chasing~\cite{ioannidis2007exploratory}.
\end{inparaenum}
\end{DocumentVersionTR}
The publication bias, on the one hand, refers to the phenomenon that the outcome of a study determines the decision to publish. Hence, statistically significant positive results are more likely to be published, than null results---even if null results live up to the same scientific rigor and possibly carry more information for falsification. Furthermore, researchers might be incentivized to engage in research practices that ensure reporting of statistically significant results, introducing biases towards questionable research practices.

The winner's curse\processifversion{DocumentVersionConfernce}{, on the other hand,} refers to the phenomenon that under-powered studies tend to report more extreme effects with statistical significance, hence tend to introducing a bias in the mean effect estimates in the field.

\processifversion{DocumentVersionTR}{Significance chasing, on the other hand, considers the phenomenon of authors producing excess significant results, especially when they their results are close to a stated significance level.}

To the best of our knowledge, these questions on reliability of statistical inferences in cyber security user studies have not been systematically answered, to date. Coopamootoo and Gro{\ss}~\cite{SLR2017} offered a manual coding of syntactic completeness indicators on studies sampled in a systematic literature review (SLR) of 10 years of cyber security user studies, while also commenting on \emph{post-hoc} power estimates for a small sub-sample. Gro{\ss}~\cite{Gross2019} investigated the fidelity of statistical test reporting along with an overview of multiple-comparison corrections and the identification of computation and decision errors. 
While we chose to base our analysis on the same published SLR sample, we close the research gap by creating a sound empirical foundation to estimate effect sizes, their standard errors and confidence intervals, by establishing power simulations vs. typical effect size thresholds, \processifversion{DocumentVersionConference}{by investigating publication bias and winner's curse.}\processifversion{DocumentVersionTR}{by investigating publication bias, winner's curse, and significance chasing.}

\paragraph{Our Contributions.}
We are the first to estimate a large number ($n=431$) of heterogenous effect sizes from cyber security user studies with their confidence intervals. Based on this estimation, we are able to show that a considerable number of tests executed in the field are underpowered, leaving results in question. This holds especially for small studies which computed a large number of tests at vanishingly low power. Furthermore, we are able to show that the reported effects of underpowered studies are especially susceptible to falter under Multiple-Comparison Corrections (MCC), while adequately powered studies are robust to MCC. 

We are the first to quantify empirically that a publication bias is present in the field of cyber security user studies. We can further evidence that the field suffers from the over-estimated effect sizes at low power, the winner's curse. \processifversion{DocumentVersionTR}{We are the first to explore this field for significance chasing, finding indications of it being present especially in the presence or absence of MCC.} We conclude our study with practical and empirically grounded recommendations for researchers, reviewers and funders.

%%%% \BEGIN{GENERATED CONTENT} %%%%

\section{Background}

\subsection{Statistical Inferences and Null Hypothesis Significance Testing}
Based on a---necessarily \emph{a priori}---specified null hypothesis (and alternative hypothesis) and a given significance level $\alpha$, statistical inference with \emph{null hypothesis significance testing}~\cite[pp. 163]{Howell2014Fundamental} sets out to establish how surprising an obtained observation $D$ is, assuming the null hypothesis being true. This is facilitated by means of a \emph{test statistic} that relates observations to appropriate probability distributions.
It is inherent to the method that the statistical hypothesis \emph{must} be fixed, before the sample is examined. 

The $p$-\emph{value}, then, is the likelihood of obtaining an observation as extreme as or more extreme than $D$, contingent on the null hypothesis being true, all assumption of the test statistic being fulfilled, the sample being drawn randomly, etc. 
Indeed, not heeding the assumptions of the test statistic is one of the more subtle ways how the process can fail.
\processifversion{DocumentVersionTR}{Null hypothesis significance testing and $p$-values are often subject to fallacies, well documented by Nickerson~\cite{Nickerson2000null}.}

Statistical inferences carry the likelihood of a false positive or \emph{Type I} error~\cite[pp. 168]{Howell2014Fundamental}. They are impacted, hence, by \emph{multiplicity}, that is, the phenomenon that computing multiple statistical tests on a test family inflates the family-wise error rate. To mitigate this effect, it is prudent practice to employ \emph{multiple-comparison corrections} (MCC)~\cite[pp. 415]{Howell2014Fundamental}. The Bonferroni correction we use here is the most conservative one, adjusting the significance level $\alpha$ by dividing it by the number of tests computed in the test family.

\subsection{Effect Sizes and Confidence Intervals}
We briefly introduce estimation theory~\cite{cumming2013understanding} as a complement to significance testing and as a key tool for this study. An observed \emph{effect size} (ES) is a \emph{point estimate} of the magnitude of an observed effect. Its \emph{confidence interval} (CI) is the corresponding interval estimate~\cite[pp. 313]{Howell2014Fundamental}. For instance, if we consider the popular $95\%$ confidence interval on an effect size, it indicates that if an experiment were repeated infinitely many times, we would expect that the point estimate on the population effect were within the respective confidence interval $95\%$ of the cases. The \emph{standard error} of an ES is equally a measure of the effects uncertainty and monotonously related to the width of the corresponding confidence interval.

Notably, confidence intervals are often misused or misinterpreted~\cite{hoekstra2014robust,morey2016fallacy}. For instance, they do \emph{not} assert that the population effect is within a point estimate's CI with $95\%$ likelihood.

However, used correctly, effect sizes and their confidence intervals are useful in establishing the practical relevance of and confidence in an effect~\cite{gardner1986confidence}. They are, thereby, recommended as minimum requirement for standard reporting\processifversion{DocumentVersionTR}{~\cite{cumming2012statistical}}, such as by the APA guidelines\processifversion{DocumentVersionConference}{~\cite{APAGuidelines6th2009}}\processifversion{DocumentVersionTR}{~\cite{APAGuidelines6th2009,publications2008reporting}}.
Whereas a statement of statistical significance or $p$-value largely gives a binary answer, an effect size quantifies the effect observed and, thereby, indicates what its impact in practice might be.

\subsection{Statistical Power}
In simple terms, \emph{statistical power} ($1 - \beta$)~\cite{Cohen1992power} is the probability that a test correctly rejects the null hypothesis, if the null hypothesis is false in reality.
Hence, power is the likelihood not to commit a false negative or \emph{Type II} error.

It should not go unnoticed that power also has an impact on the probability whether a positively reported result is actually true in reality, often referred to as \emph{Positive Predictive Value (PPV)}~\cite{ioannidis2005most}.
The lower the statistical power, the less likely a positive report is true in reality. 
Hence, a field affected by predominately low power is said to suffer from a \emph{power failure}~\cite{button2013power}.

Statistical power is largely determined by significance level, sample size, and the population effect size $\theta$.
\emph{A priori} statistical power of a test statistic is estimated by a power analysis~\cite[pp. 372]{Howell2014Fundamental} on the sample size employed vis-{\`a}-vis of the anticipated effect size, given a specified significance level $\alpha$ and target power $1-\beta$.

\emph{Post-hoc} statistical power~\cite[p. 391]{Howell2014Fundamental}, that is, computed on observed effect sizes after the face, is not only considered redundant to the $p$-value and confidence intervals on the effect sizes, but also cautioned against as treacherously misleading: It tends to overestimate the statistical power because it discounts the power lost in the study execution and because it is vulnerable to being inflated by over-estimated observed effect sizes. Hence, especially small under-powered studies with erratic effect size estimates tend to yield a biased post-hoc power. 
Hence, post-hoc power statements are best disregarded.

We offer a less biased alternative approach in \emph{power simulation}.
In that, we specify standard effect size thresholds, that is, we parametrize the analysis on assumed average effect sizes found in a field.
We then compute the statistical power of the studies given on their reported sample size against those thresholds.
As the true average effect sizes of our field are unknown, we offer power simulations for a range of typical effect size thresholds.

\subsection{Research Biases}
\label{sec:biases}
Naturally, even well-executed studies can be impacted by a range of biases on per-study level.
\begin{DocumentVersionTR}A cursory spotlighting of biases might include:
\begin{inparaenum}[(i)]
  \item sampling bias, from inappropriately selected sampling frames and non-random sampling methods,
  \item systematic and order bias, from flawed or non-random assignment,
  \item measurement bias, from flawed instruments and their application,
  \item experimenter and confirmation bias, from insufficient blinding, and
  \item a whole host of biases from poor study and statistical analysis execution as well as questionable research practices.
\end{inparaenum}
\end{DocumentVersionTR}
In this study, we consider biases of a field, instead. 
\begin{DocumentVersionConference}
We zero in on two biases, specifically:
\begin{inparaenum}[(i)]
  \item the publication bias and
  \item the winner's curse.
\end{inparaenum}
\end{DocumentVersionConference}
\begin{DocumentVersionTR}
We consider three biases, specifically:
\begin{inparaenum}[(i)]
  \item the publication bias,
  \item the winner's curse, and
  \item significance chasing.
\end{inparaenum}
\end{DocumentVersionTR}

The \emph{publication bias}~\cite{rosenthal1979file,light1984summing,dickersin1990existence,sterling1995publication} refers to the phenomenon that the publication of studies may be contingent on their positive results, hence condemning null and unfavorable results to the file-drawer~\cite{rosenthal1979file,scargle1999publication}.

The \emph{winner's curse}~\cite{button2013power} is a specific kind of publication bias referring to the phenomenon that low-power studies only reach statistically significant results on large effects and, thereby, tend to overestimate the observed effect sizes. They, hence, perpetuate inflated effect estimates in the field.

We chose them as lens for this paper because they both operate on the interrelation between sample size (impacting standard error and power) and effects observed and emphasize different aspects of the overall phenomenon. The publication bias is typically visualized with funnel plots~\cite{light1984summing}, which pit observed effect sizes against their standard errors. We shall follow Sterne and Egger's suggestion~\cite{sterne2001funnel} on using log odds ratios as best suited $x$-axis.
If no publication bias were present, funnel plots would be symmetrical. Hence, an asymmetry is an indication of bias.
This asymmetry is tested with the non-parametric rank correlation coefficient Kendall's $\tau$~\cite{begg1994operating}.
We note that funnel-plots as analysis tools can be impacted by the heterogeneity of the effect sizes investigated~\cite{terrin2003adjusting} and, hence, need to be taken with a grain of salt.

\begin{DocumentVersionTR}
\emph{Significance chasing}~\cite{ioannidis2007exploratory} refers to the presence of an excess of statistically significant results especially around a threshold significance level, such as $\alpha = .05$. 
The excess of statistically significant results is established as follows:
\[ A := \left[\frac{(O-E)^2}{E} + \frac{(O-E)^2}{(n-E)}\right], \]
where $A$ is approximated by a $chi^2$ distribution with $\vari{df}=1$, $O$ is the number of observed significant results, $n$ is the overall number of tests, and
\[ E = \sum_{i=1}^{n} (1-\beta_i). \]
For a scenario in which a sound estimate of the population effect $\hat{\theta}$ is unknown, Ioannidis and Trikalinos~\cite{ioannidis2007exploratory} proposed to evaluate the significance chasing over a range of effect size thresholds, which we shall adopt here.
Significance chasing is shown graphically by plotting the $p$-value of the significance chasing $\chi^2$-test by the significance level $\alpha$.
Apart from the research question not having been registered, the analysis is exploratory because of the nature of the method itself as explained by Ioannidis and Trikalinos and because our sample exhibits considerable heterogeneity.
\end{DocumentVersionTR}

%%%%% \END{GENERATED CONTENT} %%%%%

\section{Related Works}
\subsection{Appraisal of the Field}
Usable security, socio-technical aspects in security, human dimensions of cyber security and evidence-based methods of security are all young fields. The Systematic Literature Review (SLR) by Coopamootoo and Gro{\ss}~\cite{SLR2017}, hence, zeroed in on cyber security user studies published in the 10 years 2006--2016. The field has undergone some appraisal and self-reflection. The mentioned Coopamootoo-Gro{\ss} SLR considered completeness indicators for statistical inference, syntactically codeable from a study's reporting~\cite{coopamootoo2017CIcodebook}. These were subsequently described in a reporting toolset~\cite{coopamootoo2017cyber}. The authors found appreciable weaknesses in the field, even if there were cases of studies excelling in their rigor.
Operating from the same SLR sample, Gro{\ss}~\cite{Gross2019,Gross2019a} investigated the fidelity of statistical reporting, on completeness of the reports as well as the correctness of the reported $p$-values, finding computation and decision errors in published works relatively stable over time with minor differences between venues.

\subsection{Guidelines}
Over the timeframe covered by the aforementioned SLR, a number of authors offered recommendations for dependable, rigorous experimentation pertaining to this study.
\begin{inparaenum}[]
\item Peisert and Bishop~\cite{peisert2007design} considered the scientific design of security experiments.
\item Maxion~\cite{maxion2011making} discussed dependable experimentation, summarizing classical features of sound experiment design.
\item Schechter~\cite{schechter2013common} spoke from experience in the SOUPS program committee, offering recommendations for authors. His considerations on multiple-comparison corrections and adherence to statistical assumptions foreshadow recommendations we will make.
\item Coopamootoo and Gro{\ss}~\cite{CooGro2016} summarized research methodology, largely focusing on quantitative and evidence-based methods, discussing null hypothesis significance testing, effect size estimation, and statistical power, among other things.
\end{inparaenum}

\section{Aims}

\paragraph{Effect Sizes.}
As a stepping stone, we intend to estimate \emph{observed} effect sizes and their standard errors in a standardized format (log odds ratios).
\begin{researchquestion}[Effect Sizes and their Confidence]
What is the distribution of observed effect sizes and their 95\% confidence intervals?
How are the confidence intervals affected by multiple-comparison corrections?
\end{researchquestion}

\begin{compactdesc}
  \item[$H_{\mathsf{mcc}, 0}$:] The marginal proportions of tests' statistical significance are equal irrespective of per-study family-wise multiple-comparison corrections.
  \item[$H_{\mathsf{mcc}, 1}$:] Per-study family-wise multiple-comparison corrections impact the marginal proportions of tests' statistical significance.
\end{compactdesc}

\paragraph{Statistical Power.}
We inquire about the statistical power of studies independent from their possibly biased observed effect size estimates.
\begin{researchquestion}[Statistical Power]
What is the distribution of statistical power vis-{\`a}-vis parameterized effect size thresholds? As an upper bound achievable with given sample sizes as well as for the actual tests employed?
\end{researchquestion}
Given the unreliability of post-hoc power analysis, we pit the sample sizes employed by the studies and individual tests against the small, medium, and large effect size thresholds according to Cohen~\cite{Cohen1992power}. The actual thresholds will differ depending on the type of the effect size.

\paragraph{Publication Bias.}
We intend to inspect the relation between effect sizes and standard errors with funnel plots~\cite{light1984summing}, asking the question:
\begin{researchquestion}[Publication Bias]
To what extent does the field exhibit signs of publication bias measured in terms of relation between effect sizes and standard errors as well as asymmetry?
\end{researchquestion}

We can test statistically for the presence of asymmetry~\cite{begg1994operating} as indicator of publication bias, yielding the following hypotheses:
\begin{compactdesc}
  \item[$H_{\mathsf{bias}, 0}$:] There is no asymmetry measured as rank correlation between effect sizes and their standard errors.
  \item[$H_{\mathsf{bias}, 1}$:] There an asymmetry measured as rank correlation between effect sizes and their standard errors.
\end{compactdesc}

\paragraph{The Winner's Curse.}
We are interested whether low-powered studies exhibit inflated effect sizes and ask:
\begin{researchquestion}[Winner's Curse]
What is the relation between simulated statistical power (only dependent on group sizes) and observed effect sizes?
\end{researchquestion}

\begin{compactdesc}
  \item[$H_{\mathsf{wc}, 0}$:] Simulated power and observed effect size are independent.
  \item[$H_{\mathsf{wc}, 1}$:] There is a negative correlation between simulated power and observed effect size.
\end{compactdesc}
Hence, the winner's curse is evidenced iff greater power correlates with smaller effects.

\begin{DocumentVersionTR}
\paragraph{Significance Chasing.}
As an exploratory analysis not pre-registered, we will evaluate the presence of a phenomenon Ioannidis and Trikalinos~\cite{ioannidis2007exploratory} called ``significance chasing.''
\begin{researchquestion}[Significance Chasing]
To what extent is there an excess of significant results? To what extent is this excess pronounced around a significance level of $\alpha = .05$?
\end{researchquestion}
The former question will be tested with the statistical hypothesis:
\begin{compactdesc}
  \item[$H_{\mathsf{esr}, 0}$:] The observed number $O$ of ``positive'' results is equal to the expected number of significant results $E = \sum_{i=1}^{n} (1-\beta_i)$.
  \item[$H_{\mathsf{esr}, 1}$:] The observed number $O$ of ``positive'' results is different from the expected number $E$.
\end{compactdesc}
Due to the low power of the test, we follow the recommendation~\cite{ioannidis2007exploratory} to test these hypotheses with a significance level of $\alpha_{\mathsf{esr}} = .10$.
We will answer this question with and without multiple-comparison corrections.
\end{DocumentVersionTR}

\section{Method}
This study was registered  on the Open Science Framework (OSF)\footnote{\url{https://osf.io/bcyte/}}, before its statistical inferences commenced. Computations of statistics, graphs and tables are done in \textsf{R} with the packages \textsf{statcheck}, \textsf{metafor}, \textsf{esc}, \textsf{compute.es}, \textsf{pwr}. Their results are woven into this report with \textsf{knitr}. 
Statistics are computed as two-tailed with $\alpha =.05$ as reference significance level. Multiple-comparison corrections are computed with the Bonferroni method, adjusting the significance level used with the number of members of the test family.

\subsection{Sample}
The sample for this study is based on a 2016/17 Systematic Literature Review (SLR) conducted by Coopamootoo and Gro{\ss}~\cite{SLR2017}. This underlying SLR, its search, inclusion and exclusion criteria are reported in short form by Gro{\ss}~\cite{Gross2019} are included in this study's OSF Repository. We have chosen this SLR on the one hand, because its search strategy, inclusion and exclusion criteria are explicitly documented supporting its reproducibility and representativeness; the list of included papers is published. On the other hand, we have chosen it as sample, because there have been related analyses on qualitatively coded completeness indicators as well as statistical reporting fidelity~\cite{Gross2019} already. Therefore, we extend a common touchstone for the field.
The overall SLR sample included $N=146$ cyber security user studies. Therein, Gro{\ss}~\cite{Gross2019} identified $112$ studies with valid statistical reporting in the form of triplets of test statistic, degrees of freedom, and $p$-value. In this study, we extract effect sizes for $t$-, $\chi^2$-, $r$-, one-way $F$-tests, and $Z$-tests, complementing automated with manual extraction.

\subsection{Procedure}
\begin{figure*}
  \centering
  \includegraphics[keepaspectratio,width=\textwidth]{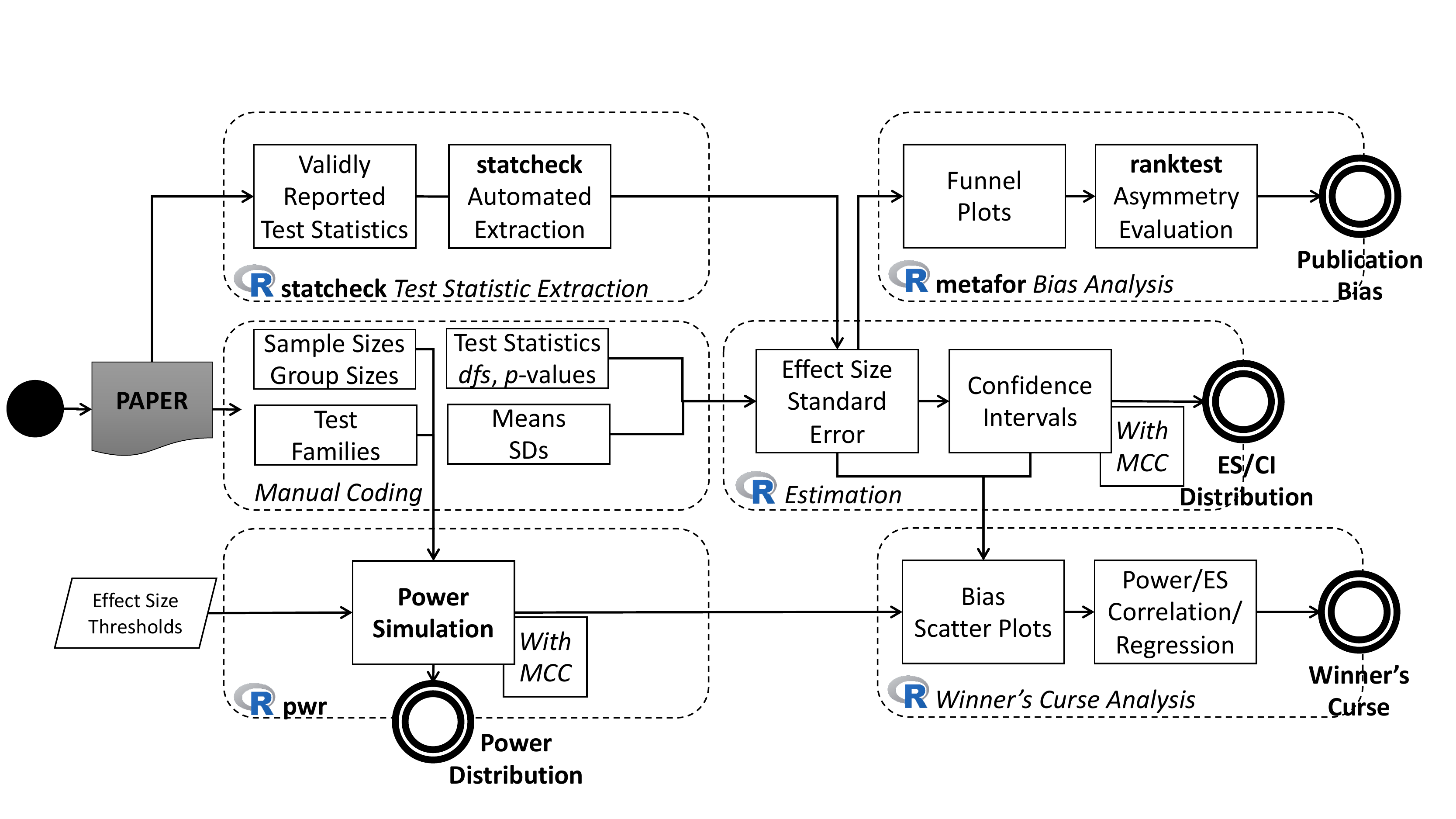}
  \caption{Flow chart of the analysis procedure}
  \label{fig:procedure}
\end{figure*}

We outlined the overall procedure in Figure~\ref{fig:procedure} and will describe the analysis stages depicted in dashed rounded boxes in turn. Figure~\ref{fig:dependency} in Appendix~\ref{app:dep} illustrates how sampled and parametrized data relate to the different analyses.

\paragraph{Automated Test Statistic Extraction.}
We analyzed the SLR sample with \textsf{R} package \textsf{statcheck} proposed by Epskamp and Nuijten~\cite{epskamp2014statcheck}. We obtained cases on test statistic type, degrees of freedom, value of the test statistic and $p$-value along with a correctness analysis. This extraction covered correctly reported test statistics (by APA guidelines) and $t$-, $\chi^2$-, $r$-, $F$-tests, and $Z$-tests at that.

\paragraph{Manual Coding.}
For all papers in the SLR, we coded the overall sample size, use of Amazon Mechanical Turk as sampling platform, and the presence of multiple-comparison corrections.
For each statistical test, we also coded group sizes, test statistics, degrees of freedom, $p$-values, means and standard deviations if applicable as well as test families. 
For the coding of test families, we distinguished different studies reported in papers, test types as well as dimensions investigated.
\processifversion{DocumentVersionTR}{For instance, if a study computed multiple One-Way ANOVAs on the same sample across the same dimensions but possibly on different variables, we would consider that a test family. We would consider pair-wise post-hoc comparisons a separate test family.}

\paragraph{Test Exclusion.} To put all effect sizes on even footing, we excluded tests violating assumption and ones not constituting one-way comparisons.

\paragraph{Power Simulation.}
\label{sec:method.power.sim}
We conducted a power simulation, that is, we specified effect size thresholds for various effect size types according to the classes proposed by Cohen~\cite{cohen1992statistical}. Table~\ref{tab:thresholds} summarizes corresponding thresholds.

% latex table generated in R 3.6.2 by xtable 1.8-4 package
% Tue Jul 14 17:52:27 2020
\begin{DocumentVersionConference}
\begin{table}[tb]
\centering
\caption{Effect size thresholds for various statistics and effect size (ES) types~\cite{Cohen1992power}}
\label{tab:thresholds}
\begin{tabular}{lcrrr}
  \toprule
\multirow{2}{*}{ES Type} &  \multirow{2}{*}{Statistic} & \multicolumn{3}{c}{Threshold}\\
\cmidrule(lr){3-5}
                                                                              && Small & Medium & Large \\ 
  \midrule
Cohen's $d$ & $t$ & 0.20 & 0.50 & 0.80 \\ 
Pearson's $r$ & $r$ & 0.10 & 0.30 & 0.50 \\ 
Cohen's $w$ & $\chi^2$ & 0.10 & 0.30 & 0.50 \\ 
Cohen's $f$ & $F$ & 0.10 & 0.25 & 0.40 \\
   \bottomrule
\end{tabular}
\end{table}
\end{DocumentVersionConference}

\begin{DocumentVersionTR}
\begin{table}[tb]
\centering
\caption{Effect size thresholds for various statistics and effect size (ES) types~\cite{Cohen1992power}}
\label{tab:thresholds}
\begin{tabular}{lcrrr}
  \toprule
\multirow{2}{*}{ES Type} &  \multirow{2}{*}{Statistic} & \multicolumn{3}{c}{Threshold}\\
\cmidrule(lr){3-5}
                                                                              && Small & Medium & Large \\ 
  \midrule
Cohen's $d$ & $t$ & 0.20 & 0.50 & 0.80 \\ 
Pearson's $r$ & $r$ & 0.10 & 0.30 & 0.50 \\ 
Cohen's $w$ & $\chi^2$ & 0.10 & 0.30 & 0.50 \\ 
Cohen's $f$ & $F$ & 0.10 & 0.25 & 0.40 \\
$\mathsf{log}(\mathit{OR})$ &  & 0.36 & 0.91 & 1.45 \\
   \bottomrule
\end{tabular}
\end{table}
\end{DocumentVersionTR}

Given the sample sizes obtained in the coding, we then computed the \emph{a priori} power against those thresholds with the \textsf{R} package \textsf{pwr}, which is independent from possible over-estimation of observed effect sizes. 
We further computed power analyses based on group sizes for reported tests, including a power adjusted for multiple comparisons in studies' test families with a Bonferroni correction. We reported those analyses per test statistic type.

\paragraph{Estimation.}
We computed a systematic estimation of observed effect sizes, their standard errors and confidence intervals.
This estimation was either based on test statistics, their degrees of freedom and group sizes used for the test or on summary statistics such as reported means, standard deviations and group sizes. We conducted the estimation with the \textsf{R} packages \textsf{esc} and \textsf{compute.es} for cases in which only test statistics were available and with the package \textsf{metafor} if we worked with summary statistics (e.g., means and standard deviations). As part of this estimation stage, we also estimated $95\%$ confidence intervals (with and without multiple-comparison corrections).

\paragraph{Publication Bias Analysis.}
We used the \textsf{R} package \textsf{metafor} to compute analyses on the publication bias.
In particular, we produced funnel plots on effect sizes and their standard errors~\cite{light1984summing}.
For this analysis, we converted all effect sizes and standard errors irrespective of their origin to log odds ratios as the predominant effect-size form for funnel plots~\cite{sterne2001funnel}.
Following the method of Begg and Mazumdar~\cite{begg1994operating}, we evaluated a rank correlation test to ascertain the presence of asymmetry.

\paragraph{Winner's Curse Analysis.}
To analyze for the winner's curse, we created scatterplots that pitted the simulated power of reported tests against the observed effect sizes extracted from the papers. We applied a Loess smoothing to illustrate the bias in the distribution.
Finally, we computed a Kendall's $\tau$ rank correlation to show the relationship between absolute effect size and power.
We employed a robust linear regression of the \textsf{R} package \textsf{MASS} using an iterated re-weighted least squares (IWLS) fitting to estimate the mean expected effect size of the field at $100\%$ power.

\begin{DocumentVersionTR}
\paragraph{Significance Chasing Analysis.}
We analyzed for significance chasing by computing the expected number of significant results $E$ from the simulated power $1-\beta$ of the respective tests against specified effect size thresholds defined in Table~\ref{tab:thresholds}. The observed number of ``positive'' results was drawn from the significance of the effects with respect to a parametrized significance level $\alpha$. We computed the $\chi^2$ significance chasing test with $\vari{df}=1$ and $\alpha_{\mathsf{esr}}$ as defined in Section~\ref{sec:biases}.
\end{DocumentVersionTR}

%%%% \BEGIN{GENERATED CONTENT} %%%%

\section{Results}

\subsection{Sample}
The sample was refined in multiple stages, first establishing papers that are candidates for effect size extraction, their refinement shown in Table~\ref{tab:sample}\processifversion{DocumentVersionConference}{ in Appendix~\ref{app:sample}}.
In total, we retained a sample of $N_{\const{studies}} = 54$ studies suitable for effect size extraction.

Secondly, we set out to extract test statistics and effect sizes with \textsf{statcheck} and manual coding.
Table~\ref{tab:sampleTests} \processifversion{DocumentVersionConference}{in Appendix~\ref{app:sample}} gives an overview how these extracted tests were first composed and then pruned in an exclusions process focused on statistical validity.
After exclusion of tests that would not yield valid effect sizes, we $N_{\const{es}} = 454$ of usable effect sizes and their standard errors.

\begin{DocumentVersionTR}
\begin{table}
\centering
\caption{Sample Refinement on Papers}
\label{tab:sample}
\begin{tabular}{lrr}
\toprule
\textbf{Phase}  & Excluded & Retained\\
\midrule
\textit{Source SLR}~\cite{SLR2017}\\
\quad Search results (Google Scholar) &    ---         & 1157\\
\quad Inclusion/Exclusion             & 1011 & 146\\
\midrule
\textit{Refinement in this study}\\
\quad Empirical studies      & 2 & 144\\
\quad With sample sizes      & 21 & 123\\
\quad With extractable tests & 69 & 54\\
\bottomrule
\end{tabular}
\end{table}

\sampleTests
\end{DocumentVersionTR}

We include the descriptives of the complete sample of extracted effect sizes grouped by their tests in Table~\ref{tab:descExtractedES}\processifversion{DocumentVersionConference}{ of Appendix~\ref{app:sample}}.
The table standardizes all effect sizes as log odds ratios, irrespective of test statistic of origin.

\begin{DocumentVersionTR}
\descExtractedES
\end{DocumentVersionTR}
\fixme{Check for negative values in descExtractedES table. Meant to be on absolute effect sizes. log odds negative because so small?}

\subsection{Effect Size Estimates and their Confidence}
\label{sec:es_ci}

In Figure~\ref{fig:caterpillarES}, we analyze the effect size estimates of our sample with their confidence intervals in a caterpillar plot: estimated effect sizes are plotted with error bars representing their $95\%$ confidence intervals and ordered by effect size. 
Two thirds of the observed effects did not pass the medium threshold: 
\begin{inparaenum}[(i)]
  \item $37\%$ were trivial, 
  \item $28\%$ were small,
  \item $15\%$ were medium, and
  \item $20\%$ were large.
\end{inparaenum}

The figure underlays the uncorrected confidence intervals (gray) with the multiple-comparison-corrected confidence intervals in red. While $54\%$ of $431$ tests were statistically significant without MCC, only $38\%$ were significant after appropriate MCC were applied. \processifversion{DocumentVersionTR}{Table~\ref{tab:descSigMCCcombined} contains the corresponding contingency tables.}
The multiple-comparison correction significantly impacted the significance of the tests, FET $p < .001$, $\vari{OR} = 0.53$, 95\% CI $[0.4, 0.7]$
We, thereby, reject the null hypothesis $H_{\mathsf{mcc}, 0}$.

\caterpillarES

\subsection{Upper Bounds of Statistical Power}
\label{sec:power.ub}

We estimate the upper-bound statistical power studies can achieve had they used their entire sample for a single two-tailed independent-samples $t$-test versus a given standardized mean difference effect size. Thereby, Figure~\ref{fig:powerSimCombinedUB} offers us a first characterization of the field in a beaded monotonously growing power plot. \processifversion{DocumentVersionTR}{The corresponding descriptive statistics are included in Table~\ref{tab:descPowerSimCombined} in Appendix~\ref{app:desc}.}

\powerSimCombinedUB

Let us unpack what we can learn from the graph. 
Regarding the sample size density on the top of Figure~\ref{fig:powerSimUB}, we observe that the sample sizes are heavily biased towards small samples ($N < 100$).
Considering the ridge power plot in Figure~\ref{fig:powerSimRidgeUB}, the middle ridge of power versus medium effects shows the field to be bipartite:
There is there is a peak of studies achieving greater than $80\%$ power against a medium effect. Those studies match the profile of studies with \emph{a priori} power analysis seeking to achieve the recommended $80\%$ power. However, roughly the same density mass is in smaller studies failing this goal.
The bottom ridge line tells us that almost no studies achieve recommended power against small effects.

\subsection{Power of Actual Tests}
\label{sec:power.actual}

\powerDensityMCC

Figure~\ref{fig:powerDensityMCC} illustrates the power distribution of all tests and all ES thresholds investigated taking into account their respective group sizes, comparing between scenarios with and without MCC.
Notably, the studies tests designed to have $80\%$ power largely retain their power under MCC. We observe a considerable number of tests with power of approx. $50\%$\processifversion{DocumentVersionTR}{, that is, just statistically significant,} which falter under MCC.
\processifversion{DocumentVersionTR}{The descriptive statistics Table~\ref{tab:descPowerSimESCombined} in Appendix~\ref{app:desc} confirms this observation.}
\begin{DocumentVersionTR}
In Figure~\ref{fig:powerDensityMCCthresholds}, we compare the power distribution by effect size thresholds and MCC.

\powerDensityMCCthresholds
\end{DocumentVersionTR}

\processifversion{DocumentVersionConferene}{Distinguishing further between different test types, we consider independent-samples $t$- and $2 \times 2$ $\chi^2$-tests as the most prevalent test statistics. Their respective power simulations are included in Figures~\ref{fig:powerSimCombinedT} and~\ref{fig:powerSimCombinedChisq} in Appendix~\ref{app:tests}, respectively.}
\processifversion{DocumentVersionTR}{Distinguishing further between different test types, we consider independent-samples $t$-, $2 \times 2$ $\chi^2$-tests, and one-way $F$-tests as test statistics. Their respective power simulations are included in Figures~\ref{fig:powerSimCombinedT}, \ref{fig:powerSimCombinedChisq}, and~\ref{fig:powerSimCombinedF} in Appendix~\ref{app:tests}, respectively.}
In both cases, we observe the following phenomena:
\begin{inparaenum}[(i)]
  \item The density mass is on smaller sample sizes.
  \item The ridge-density plots show characteristic ``two-humped'' shapes, in exhibiting a peak above $80\%$ power, but also a density mass at considerably lower power.
  \item \processifversion{DocumentVersionConference}{Both $t$-tests and $\chi^2$-tests were largely ill-equipped to detect small effect sizes.}\processifversion{DocumentVersionTR}{The three classes of $t$-, $\chi^2$-, and $F$-tests were largely ill-equipped to detect small effect sizes.}
\end{inparaenum}
Overall, we see a self-similarity of the MCC-corrected power of actual tests vis-{\`a}-vis of the upper-bound power considered in the preceding section.

\subsection{Publication Bias}
\label{sec:pub_bias}

\combinedFunnelPlots

The funnel plots in Figure~\ref{fig:combinedFunnelPlots} shows the results for $47$ papers and a total of $431$ statistical tests. For the aggregated plot~\ref{fig:funnelPlotAvg}, we computed the mean log odds ratio and mean standard error per paper.
We observe in both plots that with greater standard errors (that is, smaller samples), the effect sizes become more extreme. Hence, we conjecture that smaller studies which did not find significant effects were not published.

By the Begg-Mazumdar rank-correlation test~\cite{begg1994operating}, there is a statistically significant asymmetry showing the publication bias in the per-paper aggregate, Kendall's $\tau(N = 47) = .349$, $p < .001$., Pearson's $r = .52$, 95\% CI $[.52, .52]$.
We reject null hypothesis $H_{\mathsf{bias}, 0}$.

\subsection{The Winner's Curse}
\label{sec:wc}

\winnersCurse

In Figure~\ref{fig:winnersCurse} depicts the winner's curse phenomenon by pitting the simulated power against a threshold medium effect against the observed effect sizes. We observe that at low power, extreme results were more prevalent. At high power, the results were largely clustered closely around the predicted mean log odds.

There was a statistically significant negative correlation between power and observed effect size, that is with increasing power the observed effect sizes decrease, Kendall's $\tau(N = 396) = -.338$, $p < .001$, corresponding to an ES of Pearson's $r = -.51$, 95\% CI $[-.51, -.51]$ using Kendall's estimate\processifversion{DocumentVersionTR}{, a large effect explaining $R^2_\tau = .11$ of the variance}. We reject the winner's curse null hypothesis $H_{\mathsf{wc},0}$.

We evaluated an iterated re-weighted least squares (IWLS) robust linear regression (RLM) on the $\vari{ES} \sim \vari{power}$ relation mitigating for outliers, statistically significant at $F(1, 394) = 114.135$, $p < .001$.
We obtained an intercept of 1.6 95\% CI [1.44, 1.76], $F(1, 394) = 331.619$, $p < .001$. For every $10\%$ of power, the measured effect size decreased by -0.11; 95\% CI [-0.13, -0.09], $F(1, 394) = 114.135$, $p < .001$. 
The simulated-power regression explained approximately $R^2 = .08$ of the variance; the standard error of the regression was $S = 0.19$. \processifversion{DocumentVersionTR}{Figure~\ref{fig:winnersCurseResidual} shows the regression-residuals plot.}

\processifversion{DocumentVersionTR}{\winnersCurseResidual}

We can extrapolate to the expected mean log odds ratio at $100\%$ power $\const{log}(\vari{OR}) = 0.47$, 95\% CI $[0.21, 0.72]$. This corresponds to an SMD estimate in Cohen's $d = 0.26$, 95\% CI $[0.12, 0.4]$.

\begin{DocumentVersionTR}
\subsection{Significance Chasing}
\label{sec:sc}

By the significance-chasing $\chi^2$-test of Ioannidis and Trikalinos~\cite{ioannidis2007exploratory} for fixed $\alpha = .05$, we found a statistically significant excess of significant findings when considering the effects without multiple-comparison corrections (MCCs), for small, medium and large effect thresholds, with $\chi^2(1) = 41.791, p < .001$, $\chi^2(1) = 9.638, p = .002$, and $\chi^2(1) = 15.973, p < .001$, respectively. Consequently, we reject the null hypothesis $H_{\mathsf{esr}, 0}$.

While not present for the medium effect threshold under MCCs, $\chi^2(1) = 0.451, p = .502$, the statistically significant excess of significant findings persisted vis-{\`a}-vis of small and large thresholds, $\chi^2(1) = 400.137, p < .001$ and $\chi^2(1) = 68.369, p < .001$, respectively.

In Figure~\ref{fig:sigChasing}, we evaluated the tested effects for significance chasing by significance level, once without MCC and once with MCC. In Figure~\ref{fig:sigChasing.NoMCC} without MCC, we found a characteristic significance-chasing dip in the $\chi^2$ $p$-value below the $\alpha = .05$ threshold. This phenomenon was highlighted by Ioannidis and Trikalinos~\cite{ioannidis2007exploratory} as an indicator of authors ``chasing significance'' by making design decisions biasing towards a significant outcome when their results are just so not significant. Figure~\ref{fig:sigChasing.MCC} with MCC does not show a comparable phenomenon, allowing for the conjecture that authors made the use of multiple-comparison corrections contingent on still getting a significant result.

\sigChasing
\end{DocumentVersionTR}

%%%%% \END{GENERATED CONTENT} %%%%%

\section{Discussion}

\paragraph{The power distribution is characteristically two-humped.}
We found empirical evidence that a substantive number of studies and half the tests extracted were adequate for $80\%$ power at a medium target effect size. Hence, it is plausible to conjecture an unspoken assumption in the field that the population effect sizes in cyber security user studies are medium (e.g., Cohen's $d \geq .50$). The good news here is that studies that were appropriately powered, that is, aiming for $80\%$ power, retained that power also under multiple-comparison corrections. Studies which were under-powered in the first place, got entangled by MCCs and ended up with negligible power retained (cf. Figure~\ref{fig:powerDensityMCC}, Section~\ref{sec:power.actual}).

Having said that, the power distribution for the upper bound as well as for actual tests (Figures~\ref{fig:powerSimCombinedUB}, \ref{fig:powerSimCombinedT}, \processifversion{DocumentVersionConference}{and~\ref{fig:powerSimCombinedChisq}}\processifversion{DocumentVersionTR}{ \ref{fig:powerSimCombinedChisq}, and~\ref{fig:powerSimCombinedF}}) came in two ``humps.''
While we consistently observed peaks at greater than $80\%$ power for medium effect sizes, there was a density mass of under-powered tests, where the distribution was roughly split half-half.
Typically, tests were altogether too under-powered to detect small effect sizes. Overall, we believe we have evidence to attest a power failure in the field.

\paragraph{Population effect sizes may be smaller than we think.}
The problem of power failure is aggravated by the mean effect sizes in the SLR having been close to small, shown in the caterpillar forest plot (Figure~\ref{fig:caterpillarES}) and the ES descriptives (Table~\ref{tab:descExtractedES}).
In fact, our winner's curse analysis estimated a mean Cohen's $d = 0.26$, 95\% CI $[0.12, 0.4]$. 
Of course, it is best to obtain precise effect size estimates for the effect in question from prior research, ideally from systematic meta-analyses deriving the estimate of population effect size $\hat{\theta}$.
Still, the low effect size indicated here should give us pause: aiming for a medium effect size as a rule of thumb might be too optimistic.

\paragraph{Cyber security user studies suffer from a host of biases.}
We showed the presence of an appreciable publication bias (cf. Figure~\ref{fig:combinedFunnelPlots}, Section~\ref{sec:pub_bias}), that is, the phenomenon that the publication of studies was contingent on their positive outcomes, and found evidence of the winner's curse, that is, the phenomenon that under-powered studies yielded exaggerated effect estimates (cf. Figure~\ref{fig:winnersCurse}, Section~\ref{sec:wc}). 

\begin{DocumentVersionTR}
Taking as a grain of salt the heterogeneity of the studies and of the exploratory nature of the method involved, we found in Section~\ref{sec:sc} that there is a statistically significant excess of significant results in cyber security user studies, as well. We have further seen indications of significance chasing around a significance level of $\alpha =.05$ which are absent when multiple-comparison corrections are taken into account.  
\end{DocumentVersionTR}

Taken together with the  likely close-to-small population effect sizes and the diagnosed power failure, we need to conclude that the field is prone to accept publications that are seemingly ``positive'' results, while perpetuating biased studies with over-estimated effect sizes. These issues could be resolved with a joint effort by field's stakeholders---authors, gatekeepers and funders: paying greater attention to statistical power, point and interval estimates of effects, and adherence to multiple-comparison corrections.

\subsection{Limitations}
\paragraph{Generalizability.} We observe that we needed to exclude a considerable number of studies and statistical tests.
This is consistent with the observations by Coopamootoo and Gro{\ss}~\cite{SLR2017} on prevalent reporting completeness, finding that $71\%$ of their SLR sample did not follow standard reporting guidelines  and only $31\%$ combinations of actual test statistic, $p$-value and corresponding descriptives. Similarly, Gro{\ss}~\cite{Gross2019} found that 69 papers ($60\%$) did not contain a single completely reported test statistic. Hence, we also observe that meta research is severely hamstringed by the reporting practices found in the field.

We note, further, that we needed to exclude $104$ extracted statistical tests and effect sizes due to problems in how these tests were employed, leading to $17$ less represented. Studies that inappropriately used independent-sample tests in a dependent-sample research designs or violated other assumptions by, e.g., using difference-between-means test statistics (expecting a $t$ distribution) to test differences between proportions ($z$-distribution), needed to be excluded to prevent perpetuation of those issues. Finally, we needed to exclude $74$ tests because papers reported tests with degrees of freedom $\vari{df} >1$ without the summary statistics to establish the effect sizes. Even though those studies contained complete reports the auxiliary data to estimate the effects were missing.

These exclusions on empirical grounds limit generalizability. The retained sample of $431$ tests is focused on the studies that were most diligent in their reporting. This fact, however, makes our investigation more conservative rather than less so.

\paragraph{This is not a meta-analysis.} Proper meta-analysis combines effect sizes on \emph{similar} constructs to summary effects. Given that studies operating on the same constructs are few and far between in cyber security user studies, we standardized all effects to log odds ratios to gain a rough overall estimate of the field.

\begin{DocumentVersionTR}
There are a number of consequences of this approach. First, this study does not attempt to create an estimate of the population effect $\hat{\theta}$ for the research questions investigates as a meta-analysis would. The extrapolation of the linear regression of effect size by simulated power only offers a rough estimate of the mean effect sizes typically present.

Second, with the only common ground of the sample is that they constitute user studies in cyber security, the analysis faces considerable heterogeneity. This heterogeneity, in turn, can impact the tests on publication bias, winner's curse, and significance chasing. While the analysis of Terrin et al.~\cite{terrin2003adjusting} cautions against the use of funnel plots in face of such heterogeneity, Ioannidis and Trikalinos~\cite{ioannidis2007exploratory} make similar observations about their significance chasing detection method. In essence, Terrin et al.~\cite{terrin2003adjusting} argue that ``an inverse relation between treatment benefit and precision would result [because of researchers' design decisions], even when there is no publication bias.'' For instance, researchers who investigate effects with larger well-assured estimates $\hat{\theta}$ or who have intense interventions with strong effect sizes at their disposal, could have made a conscious decision to work with lower---but still sufficient---sample sizes. However, in the field of cyber security user studies assured estimates of population effects, e.g., as gained from systematic meta-analyses, or highly effective validated interventions are few and far between. Hence, we believe we can rule out this alternative explanation in this case.
What will likely still apply to this work is the observation that ``low quality studies may report exaggerated treatment benefit, and may also tend to be small;''~\cite{terrin2003adjusting} we observed exaggerated effect sizes of small studies in our winner's curse analysis of Section~\ref{sec:wc}. However, needing to choose between the community being biased toward publication contingent on ``positive'' results (publication bias) or prone to accept under-powered studies with exaggerated results (winner's curse), is being stuck between a rock and a hard place.
\end{DocumentVersionTR}

\section{Concluding Recommendations}
We are the first to evaluate the statistical reliability of this field on empirical grounds. While there is a range of possible explanations of the phenomena we have found---including questionable research practices in, e.g., shirking multiple-comparison corrections in search of significant findings, missing awareness of statistical power and multiplicity, or limited resources to pursue adequately powered studies---we believe the evidence of power failure, possibly close-to-small population effect sizes, and biased findings can lead to empirically underpinned recommendations. We believe that these issues, however, are systemic in nature and that the actions of different stakeholders are, thereby, inter-dependent. Hence, in the following we aim at offering recommendations to different stakeholder, making the assumption that they aim at advancing the knowledge of the field to the best of their ability and resources.

\paragraph{Researchers.} The most important recommendation here is: \emph{plan ahead with the end in mind}. That starts with inquiring typical effect sizes for the phenomena investigated. If the reported confidence intervals thereon are wide, it is prudent to choose a conservative estimate. It is tempting to just assume a medium effect size (e.g., Cohen's $d = 0.5$) as aim, but there is no guarantee the population effect sizes are that large. Our study suggests they are not. 

While it is a prudent recommendation to conduct an \emph{a priori} power analysis, we go a step further and recommend to anticipate multiple comparisons one might make. Adjusting the target significance level with a Bonferroni correction for that multiplicity can prepare the ground for a study retaining sufficient power all the way. This kind of foresight is well supported by a practice of spelling out the research aims and intended statistical inferences \emph{a priori} (e.g., in a pre-registration). Taken together these measures aim countering the risk of a power failure.

Speaking from our experience of painstakingly extracting effects from a body of literature, we are compelled to emphasize: One of the main points of strong research is that it is \emph{reusable} by other scientists. This goal is best served by reporting effect sizes and their confidence intervals as well as full triplets of test statistic, degrees of freedom and exact $p$-values, while also offering all summary statistics to enable others to re-compute the estimates. It is worth recalling that \emph{all} tests undertaken should be reported and that rigorous, well-reported studies have intrinsic value, null result or not. This line of recommendations aims at enabling the gatekeepers of the field to do their work efficiently.

\paragraph{Gatekeepers.} It bears repeating that the main goal of science is to \emph{advance the knowledge of a field}. 
With reviewers, program chairs and editors being gatekeepers and the arbiters of this goal, it is worthwhile to consider that the goal is not served well in pursuing shiny significant results or valuing novelty above all else. Such a value system is prone to fail to ward against publication and related biases. A well-powered null result or replication attempt can go a long way in initiating the falsification of a theory in need of debunking. Because empirical epistemology is rooted in falsification and replication, we need the multiple inquiries on the same phenomena. We should strive to include adequately-powered studies of sufficient rigor irrespective of the ``positiveness'' of the results presented, exercising the cognitive restraint to counter publication bias.

Reviewers can support this by insisting on systematic reporting and on getting to see \emph{a priori} specifications of aims, research designs, tests conducted, as well as sample size determinations, hence creating an incentive to protect against power failure. This recommendation dovetails with the fact that statistical inference is contingent on fulfilling the assumptions of the tests used, where the onus of proof is with the researchers to ascertain that all assumptions were satisfied. Those recommendations are in place to enable the gatekeepers to effectively ascertain the statistical validity and reliability of studies at hand.

\paragraph{Funders.} With significant investments being made in putting cyber security user studies on an evidence-based footing, we recall: ``\emph{Money talks.}'' On the one hand, we see the responsibility with the funders to support studies with sufficient budgets to obtain adequately powered samples---not to speak of adequate sampling procedures and representativeness. 
On the other hand, the funders are in a strong position to mandate \emph{a priori} power analyses, pre-registrations, strong reporting standards geared towards subsequent research synthesis, published datasets, and open-access reports.
They could, furthermore, incentivize and support the creation registered-study databases to counter the file-drawer problem.

\section*{Acknowledgment}

We would like to thank the anonymous reviewers of STAST 2020 for their comments. 
Early aspects of this study were in parts funded by the UK Research Institute in the Science of Cyber Security (RISCS) under a National Cyber Security Centre (NCSC) grant on ``Pathways to Enhancing Evidence-Based Research Methods for Cyber Security.'' Thomas Gro{\ss} was funded by the \CASCAde.

%\balance

\bibliographystyle{splncs04}
\bibliography{methods_resources,slr_user_studies,stat_check,slr,r_tools,slr_meta,guides}

\begin{thebibliography}{10}
\providecommand{\url}[1]{\texttt{#1}}
\providecommand{\urlprefix}{URL }
\providecommand{\doi}[1]{https://doi.org/#1}

\bibitem{APAGuidelines6th2009}
{American Psychological Association} (ed.): Publication Manual of the American
  Psychological Association (6th revised ed.). American Psychological
  Association (2009)

\bibitem{begg1994operating}
Begg, C.B., Mazumdar, M.: Operating characteristics of a rank correlation test
  for publication bias. Biometrics pp. 1088--1101 (1994)

\bibitem{button2013power}
Button, K.S., Ioannidis, J.P., Mokrysz, C., Nosek, B.A., Flint, J., Robinson,
  E.S., Munaf{\`o}, M.R.: Power failure: why small sample size undermines the
  reliability of neuroscience. Nature Reviews Neuroscience  \textbf{14}(5),
  365--376 (2013)

\bibitem{Cohen1992power}
Cohen, J.: A power primer. Psychological bulletin  \textbf{112}(1), ~155 (1992)

\bibitem{cohen1992statistical}
Cohen, J.: Statistical power analysis. Current directions in psychological
  science  \textbf{1}(3),  98--101 (1992)

\bibitem{SLR2017}
Coopamootoo, K., Gro{\ss}, T.: Systematic evaluation for evidence-based methods
  in cyber security. Technical Report {TR}-1528, Newcastle University (2017)

\bibitem{CooGro2016}
Coopamootoo, K.P., Gro{\ss}, T.: Evidence-based methods for privacy and
  identity management. In: IFIP International Summer School on Privacy and
  Identity Management. pp. 105--121. Springer (2016)

\bibitem{coopamootoo2017CIcodebook}
Coopamootoo, K.P., Gro{\ss}, T.: A codebook for experimental research: The
  nifty nine indicators v1.0. Tech. Rep. {TR}-1514, Newcastle University
  (November 2017)

\bibitem{coopamootoo2017cyber}
Coopamootoo, K.P., Gro{\ss}, T.: Cyber security and privacy experiments: A
  design and reporting toolkit. In: IFIP International Summer School on Privacy
  and Identity Management. pp. 243--262. Springer (2017)

\bibitem{cumming2013understanding}
Cumming, G.: Understanding the new statistics: Effect sizes, confidence
  intervals, and meta-analysis. Routledge (2013)

\bibitem{cumming2012statistical}
Cumming, G., Fidler, F., Kalinowski, P., Lai, J.: The statistical
  recommendations of the american psychological association publication manual:
  Effect sizes, confidence intervals, and meta-analysis. Australian Journal of
  Psychology  \textbf{64}(3),  138--146 (2012)

\bibitem{dickersin1990existence}
Dickersin, K.: The existence of publication bias and risk factors for its
  occurrence. Jama  \textbf{263}(10),  1385--1389 (1990)

\bibitem{epskamp2014statcheck}
Epskamp, S., Nuijten, M.: \textsf{statcheck}: Extract statistics from articles
  and recompute $p$-values (\textsf{R} package version 1.0.0.).
  \url{https://cran.r-project.org/web/packages/statcheck/index.html} (2014)

\bibitem{gardner1986confidence}
Gardner, M.J., Altman, D.G.: Confidence intervals rather than p values:
  estimation rather than hypothesis testing. Br Med J (Clin Res Ed)
  \textbf{292}(6522),  746--750 (1986)

\bibitem{Gross2019}
Gro{\ss}, T.: Fidelity of statistical reporting in 10 years of cyber security
  user studies. In: In proceedings of the 9th International Workshop on
  Socio-Technical Aspects in Security (STAST'2019). LNCS, vol. 11739, pp.
  1--24. Springer Verlag (2019)

\bibitem{Gross2019a}
Gro{\ss}, T.: Fidelity of statistical reporting in 10 years of cyber security
  user studies [extended version]. arXiv Report arXiv:2004.06672, Newcastle
  University (2020)

\bibitem{hoekstra2014robust}
Hoekstra, R., Morey, R.D., Rouder, J.N., Wagenmakers, E.J.: Robust
  misinterpretation of confidence intervals. Psychonomic Bulletin \& Review
  \textbf{21}(5),  1157--1164 (2014)

\bibitem{Howell2014Fundamental}
Howell, D.C.: Fundamental Statistics for the Behavioral Sciences (8th
  international ed). Cengage Learning (2014)

\bibitem{ioannidis2005most}
Ioannidis, J.P.: Why most published research findings are false. PLoS Med
  \textbf{2}(8), ~e124 (2005)

\bibitem{ioannidis2007exploratory}
Ioannidis, J.P., Trikalinos, T.A.: An exploratory test for an excess of
  significant findings. Clinical trials  \textbf{4}(3),  245--253 (2007)

\bibitem{light1984summing}
Light, R.J., Pillemer, D.B.: Summing up; the science of reviewing research.
  Cambridge, MA (USA) Harvard Univ. Press (1984)

\bibitem{maxion2011making}
Maxion, R.: Making experiments dependable. In: Dependable and Historic
  Computing, pp. 344--357. Springer (2011)

\bibitem{morey2016fallacy}
Morey, R.D., Hoekstra, R., Rouder, J.N., Lee, M.D., Wagenmakers, E.J.: The
  fallacy of placing confidence in confidence intervals. Psychonomic bulletin
  \& review  \textbf{23}(1),  103--123 (2016)

\bibitem{Nickerson2000null}
Nickerson, R.S.: Null hypothesis significance testing: a review of an old and
  continuing controversy. Psychological methods  \textbf{5}(2), ~241 (2000)

\bibitem{peisert2007design}
Peisert, S., Bishop, M.: How to design computer security experiments. In: Fifth
  World Conference on Information Security Education. pp. 141--148. Springer
  (2007)

\bibitem{publications2008reporting}
{Publications, {APA} and on Journal, Communications Board Working Group}:
  Reporting standards for research in psychology: Why do we need them? what
  might they be? The American Psychologist  \textbf{63}(9), ~839 (2008)

\bibitem{rosenthal1979file}
Rosenthal, R.: The file drawer problem and tolerance for null results.
  Psychological bulletin  \textbf{86}(3), ~638 (1979)

\bibitem{scargle1999publication}
Scargle, J.D.: Publication bias (the" file-drawer problem") in scientific
  inference. arXiv preprint physics/9909033  (1999)

\bibitem{schechter2013common}
Schechter, S.: Common pitfalls in writing about security and privacy human
  subjects experiments, and how to avoid them.
  \url{https://www.microsoft.com/en-us/research/wp-content/uploads/2016/02/commonpitfalls.pdf}
  (2013)

\bibitem{sterling1995publication}
Sterling, T.D., Rosenbaum, W.L., Weinkam, J.J.: Publication decisions
  revisited: The effect of the outcome of statistical tests on the decision to
  publish and vice versa. The American Statistician  \textbf{49}(1),  108--112
  (1995)

\bibitem{sterne2001funnel}
Sterne, J.A., Egger, M.: Funnel plots for detecting bias in meta-analysis:
  guidelines on choice of axis. Journal of clinical epidemiology
  \textbf{54}(10),  1046--1055 (2001)

\bibitem{terrin2003adjusting}
Terrin, N., Schmid, C.H., Lau, J., Olkin, I.: Adjusting for publication bias in
  the presence of heterogeneity. Statistics in medicine  \textbf{22}(13),
  2113--2126 (2003)

\end{thebibliography}

\begin{appendix}
\begin{DocumentVersionTR}
\section{Underlying Systematic Literature Review}
\label{sec:SLR}

This meta-analytic study is based on a Systematic Literature Review (SLR), which was conducted in 2016/17 for the UK Research Institute in the Science of Cyber Security (RISCS). We adapted this description of the SLR's search from its technical report~\cite{SLR2017}.  It was previously published in the supplementary materials of the analysis of statistical reporting fidelity~\cite{Gross2019a}.

\subsection{Search Strategy of the SLR Sample}
\label{sec:SLR_search}
The SLR included security and privacy papers published between 2006 and 2016 (inclusive).

The search was restricted to the following security and privacy venues:
\begin{compactitem}
  \item journals: IEEE Transactions on Dependable \& Secure Computing (TDSC), ACM Transactions on Information and System Security (TISSEC),
  \item flagship security conferences: IEEE S\&P, ACM CCS, ESORICS, and PETS or
  \item specialized conferences and workshops: LASER, SOUPS, USEC and WEIS.
\end{compactitem}

The search was conducted on Google Scholar.
Each query extracts articles mentioning ``\emph{user study}'' and at least one of the words ``\emph{experiment},''
``\emph{evidence}'' or ``\emph{evidence based}.'' The described query was executed for each of the $10$ publication venues.
In the advanced search option of Google Scholar, each of the following fields were set:
\begin{compactitem}
\item with all words = \emph{user study}
\item at least one of the words = \emph{experiment evidence ``evidence based''}
\item where my words occur = \emph{anywhere in the article}
\item return articles published in = [publication venue]
\item return articles dated between = \emph{2006--2016}
\end{compactitem}

The search yielded $1157$ publications.

\subsection{SLR Inclusion/Exclusion Criteria}
\label{sec:SLR_incl-excl}
We adapt the inclusion/exclusion criteria of the 2017 SLR~\cite{SLR2017} for this pre-registration.
The SLR focused on human factors studies including a human sample.
The following \emph{Inclusion Criteria} were applied to its overall pool of $1157$ publications:
\begin{compactitem}
 \item Studies including a user study with human participants.
  \item Studies concerned with evidence-based methods or eligible for hypothesis testing and statistical inference.
  \item Studies that lend themselves to quantitative evaluation, quoting statements of statistical significance, $p$-values or effect sizes.
  \item Studies with true experiments, quasi-experiments or observational analysis.
\end{compactitem}
Of the papers included, the ones fulfilling the following \emph{Exclusion Criteria} were excluded:
\begin{compactitem}
  \item Papers that were not subject to research peer-review, key note statements, posters and workshop proposals.
  \item Position papers or informal arguments.d R
  \item Papers not including a study with human participants,
  \item Theoretical papers.
  \item Studies with qualitative methodology.
\end{compactitem}
This inclusion/exclusion process yielded a final sample of $146$ publications.
\end{DocumentVersionTR}

%%%% \BEGIN{GENERATED CONTENT} %%%%
\begin{DocumentVersionConference}
\section{Sample Characteristics}
\label{app:sample}

We offer a detailed sample refinement on papers in Figure~\ref{tab:sample} and extracted statistical tests and corresponding effect sizes in Figure~\ref{tab:sampleTests}.

\begin{table}
\centering
\caption{Sample Refinement on Papers}
\label{tab:sample}
\begin{tabular}{lrr}
\toprule
\textbf{Phase}  & Excluded & Retained\\
\midrule
\textit{Source SLR}~\cite{SLR2017}\\
\quad Search results (Google Scholar) &    ---         & 1157\\
\quad Inclusion/Exclusion             & 1011 & 146\\
\midrule
\textit{Refinement in this study}\\
\quad Empirical studies      & 2 & 144\\
\quad With sample sizes      & 21 & 123\\
\quad With extractable tests & 69 & 54\\
\bottomrule
\end{tabular}
\end{table}

\sampleTests

\descExtractedES
\end{DocumentVersionConference}

\section{Data Dependencies}
\label{app:dep}
\begin{figure}[tb]
\centering
%\vspace{-2cm}
\includegraphics[keepaspectratio, width=\textwidth]{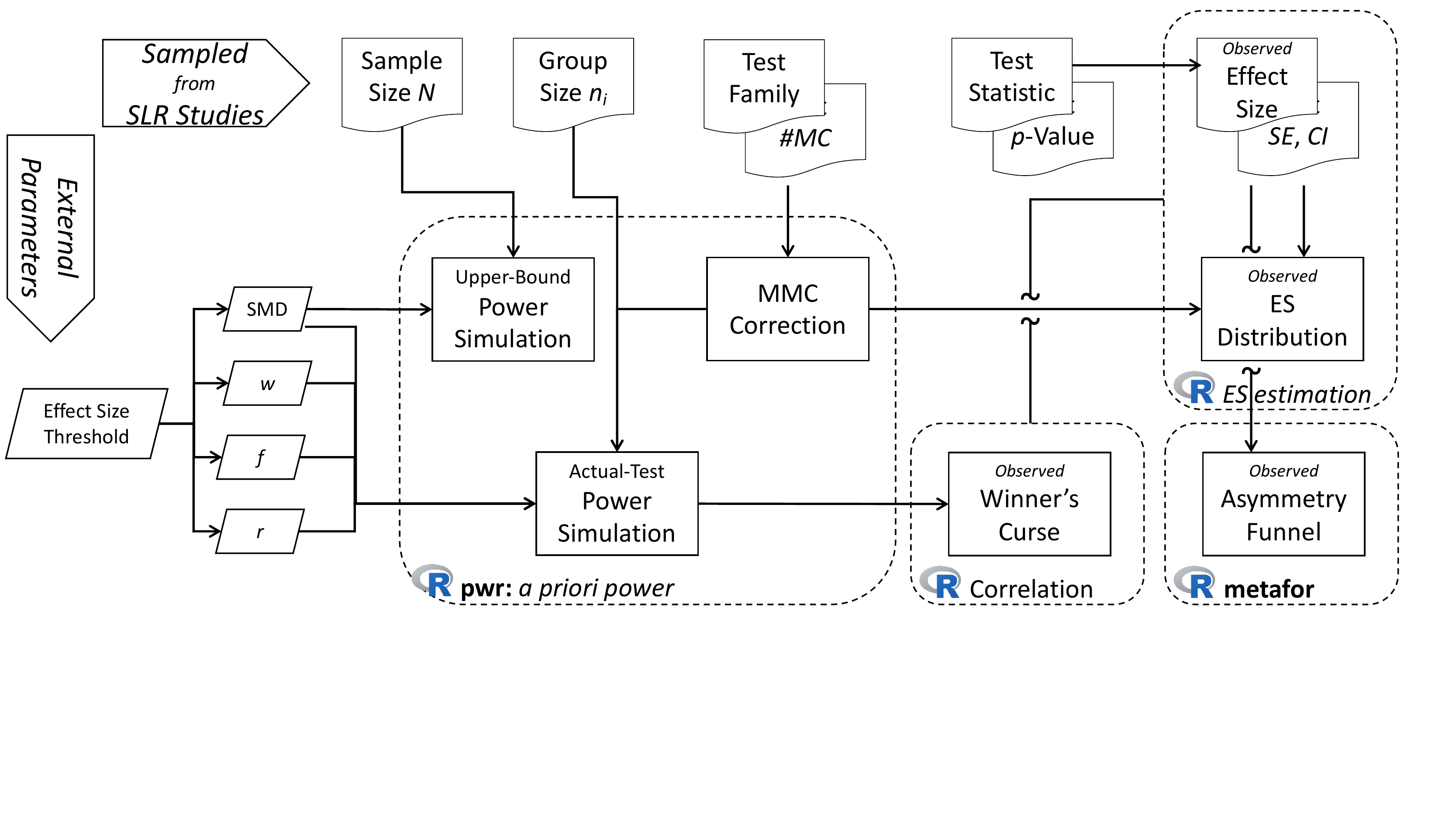}
\vspace{-2cm}
\caption{Data dependency diagram for analyses included in the paper.}
\label{fig:dependency}
\end{figure}

This work explicitly steps away from post-hoc power analysis and focuses on
statistical power simulation. Given how inflated ES estimates lead to inflated---and thereby unreliable---post-hoc power estimates, this distinction is crucial. To make clear how different analyses depend on sampled data from papers and externally inputted parameters as well as to support reproducibility, we provide a data dependency diagram in Figure~\ref{fig:dependency}.

\begin{DocumentVersionTR}
\section{Descriptive Statistics}
\label{app:desc}
We include a number of descriptive statistics complementing plots in the body of the paper. Table~\ref{tab:descSigMCCcombined} offers contingency tables on the difference made by using family-wise multiple-comparison corrections. Table~\ref{tab:descPowerSimCombined} provides the characteristics of the extracted overall sample sizes and upper-bound power simulation. Finally, Table~\ref{tab:descPowerSimESCombined} includes the power simulation for actual tests.

\descSigMCCcombined

\descPowerSimCombined

\descPowerSimESCombined
\end{DocumentVersionTR}

\section{Power of Actual Tests}
\label{app:tests}
\processifversion{DocumentVersionConference}{Figures~\ref{fig:powerSimCombinedT} and~\ref{fig:powerSimCombinedChisq} include the power estimates for actual independent-samples $t$-tests and $2 \times 2$ $\chi^2$ tests, respectively.}
\processifversion{DocumentVersionTR}{Figures~\ref{fig:powerSimCombinedT}, \ref{fig:powerSimCombinedChisq}, and~\ref{fig:powerSimCombinedF} include the power estimates for actual independent-samples $t$-tests, $2 \times 2$ $\chi^2$ tests, one-way $F$-tests, respectively.}

\powerSimCombinedT

\powerSimCombinedChisq

\begin{DocumentVersionTR}
\powerSimCombinedF
\end{DocumentVersionTR}

%%%%% \END{GENERATED CONTENT} %%%%%
\end{appendix}

\end{document}